\documentclass[twocolumn]{emulateapj}
\usepackage{color}
\usepackage{url}
\usepackage[usenames,dvipsnames,svgnames,table]{xcolor}
\usepackage[colorlinks=true,
           linkcolor=blue,
           urlcolor=blue,%
           citecolor=blue]{hyperref}

\begin{document}

\title{OBSERVATION OF A QUASI-PERIODIC PULSATION IN HARD X-RAY, RADIO AND EXTREME-ULTRAVIOLET WAVELENGTHS}

\author{PANKAJ KUMAR\altaffilmark{1}, VALERY M. NAKARIAKOV\altaffilmark{2,3,4}, KYUNG-SUK CHO\altaffilmark{1}}
\affil{$^1$Korea Astronomy and Space Science Institute (KASI), Daejeon, 305-348, Republic of Korea}
\affil{$^2$Centre for Fusion, Space and Astrophysics, Department of Physics, University of Warwick, CV4 7AL, UK}
\affil{$^3$School of Space Research, Kyung Hee University, Yongin, 446-701, Gyeonggi, Republic of Korea}
\affil{$^4$ Central Astronomical Observatory at Pulkovo of RAS, 196140 St Petersburg, Russia}
\email{pankaj@kasi.re.kr}

\begin{abstract}
We present multi-wavelength analysis of a quasi-periodic pulsation (QPP) observed in the hard X-ray, radio, and extreme-ultraviolet (EUV) channels during an M1.9 flare occurred on 23--24 September 2011. The non-thermal hard X-ray emission in 25-50 keV
observed by RHESSI shows five distinct impulsive peaks of decaying
amplitude with a period of about three minutes. Similar QPP was observed in the microwave emission recorded by the Nobeyama Radioheliograph and Polarimeter in the 8.8, 15, 17~GHz channels. Interestingly, the 3-min QPP was also observed in the metric and decimetric radio frequencies (25--180, 245, 610 MHz) as repetitive type III bursts. Multi-wavelength observations from the SDO/AIA, Hinode/SOT, and STEREO/SECCHI suggest a fan-spine topology at the eruption site, associated with the formation of a quasi-circular ribbon during the flare. A small filament was observed below the fan-loops before the flare onset. The filament rose slowly and interacted with the ambient field. This behaviour was followed by an untwisting motion of the filament. Two different structures of the filament showed $\sim$3-min periodic alternate rotation in the clockwise and counterclockwise directions.  
The 3-min QPP was found to highly correlate with 3-min oscillations in a nearby sunspot.
We suggest that the periodic reconnection (modulated either by sunspot slow-mode wave or by untwisting filament) at a magnetic null-point most likely causes the repetitive particle acceleration, generating the QPP observed in hard X-ray, microwave and type III radio bursts.       

 \end{abstract}

\section{INTRODUCTION}
Quasi-periodic pulsations (QPPs) with periods from a few seconds to several minutes are often observed during solar flares in X-ray, radio, and optical wavelength bands \citep[e.g.][]{asc2004,nakariakov2009,nakariakov2010}. Similar QPP have also been detected in stellar flares \citep[e.g.][]{mathioudakis2003,pandey2009,anfinogentov2013,balona2015,pugh2015}. 
The study of the QPP phenomenon is promising in the context of understanding the basic physics of flaring energy releases, charged particle acceleration, and for the indirect estimation of plasma parameters by the method of magnetohydrodynamic (MHD) seismology \citep[e.g.][]{roberts2000,demoortel2012,liu2014}. 

Production of QPPs in flares may be associated with several physical mechanisms associated with flares. The following mechanisms could be responsible for the QPP formation:

(i) QPP may be associated with a periodic, bursty regime of spontaneous magnetic reconnection, e.g. by tearing of the current sheet associated with the formation of multiple plasmoids \citep{kliem2000}. The coalescence and interaction of different plasmoids moving up and down along the flaring current sheet can periodically modulate the bidirectional (up and down) acceleration of electrons \citep{barta2007,barta2008}. In this case, drifting pulsating structures (DPSs) in the decimetric radio frequencies are expected to be observed \citep{karlicky2010}. Recently, multiple plasmoids moving bidirectionally along an apparent  current sheet have been observed in EUV \citep{Takasao2012}. Simultaneous observations of multiple plasmoids and drifting pulsating structures (for a duration of $\sim$1~min) have been recently reported by \citet{kumar2013p}. Therefore, tearing of the current sheet and formation of multiple plasmoids can generate QPPs in radio and hard X-ray with periods from seconds to several minutes. 

(ii) Coalescence of current-carrying coronal loops can result in periodic variation of the coalescence current sheet width, accompanied by periodic production of non-thermal electrons. It would generate QPPs in hard X-ray and radio bands \citep{tajima1987,kumar2010l}. In particular, the 25--48-s QPPs detected in the rising phase of solar flares were associated with this mechanism by \citet{farnik2003}.
 
(iii) In the model recently proposed by \citet{takasao2015}
numerical simulations of the shock formation and evolution of the thermal structure in and above post-flare loops demonstrated that the strength of the termination shock shows a quasi-periodic oscillation. This scenario can also be classified spontaneously-generated QPP.

(iv) The interaction of MHD waves (e.g., fast-mode wave, slow-mode wave or torsional Alfv\'en waves) with the reconnection site may lead to induced QPPs. \citet{ning2004} studied recurrent explosive events with a period of $\sim$3--5 min, and suggested that their triggering could be caused by an MHD wave. 
A fast-mode MHD wave model proposed by \citet{nakariakov2006} links triggering of QPP with a transverse oscillation of a coronal loop situated near the reconnection region/current sheet. 
\citet{doyle2006} studied the repetitive occurrence of explosive events at a coronal hole boundary, and suggested the periodic triggering of reconnection by a kink oscillation of the flux tubes in the closed field line region.
\citet{chen2006} modelled explosive events in the transition region, and showed that the repetitive reconnection (e.g., with the period of 3--5~min) could be triggered by a slow-mode MHD wave.
This idea was further developed by \citet{sych2009} who demonstrated that 3-min QPP observed during a flare could be triggered by the along-the-field leakage of 3-min slow-mode oscillations in the umbra of a nearby sunspot. In two-ribbon flares, repetitive reflection of flare-excited slow-mode waves from footpoints of the flaring arcade may also be a possible mechanism for triggering the periodic reconnection \citep{nakariakov2011}. In that scenario, the hard X-ray source moves with a subsonic and sub-Alfv\'enic speed along the neutral line, which is consistent with observations.

Despite significant progress in both the theoretical modelling and observational detection, the specific mechanisms for QPP observed in solar and stellar flares are still not well understood and need further investigations. In particular, analysis of QPPs  in high-resolution multi-wavelength observations (e.g., simultaneously in EUV, X-rays  and radio) can provide a more complete picture of the flare event, and allow us to draw important conclusions on the QPP generation mechanisms.  

Recently, \citet{kumar2015} studied a decaying QPP with the period about 202~s in the 6--12~keV X-ray and EUV channels. They found that the QPP was generated by periodic reconnection caused by an untwisting small filament at a 3D null point in a fan-spine magnetic topology. Energetic particles accelerated along the arcade loops toward the opposite footpoints and precipitated there, were associated with periodic EUV brightening. Apart from the 202-s QPP, a 409-s oscillation was also observed as a periodic alternate flow along the arcade loops, and interpreted as a slow-mode (longitudinal) wave excited by a flare triggered at one of the footpoints of the arcade loops.  

In this paper, we analyse a QPP observed in hard X-rays (25--50 keV), microwaves (17 GHz), and EUV emission in an M1.9 flare on 23--24 September 2011. 
The flare exhibits a quasi-circular ribbon, which suggests the possible magnetic configuration to be a fan-spine topology at the boundary of the active region NOAA 11302. In Section~2, we present the observations and results. In the last section, we summarize and discuss our results.

\section{OBSERVATIONS AND RESULTS}

The {\it Atmospheric Image Assembly} (AIA; \citealt{lemen2012}) onboard the {\it Solar Dynamics Observatory}  
(SDO; \citealt{pesnell2012}) captures full disk images of the Sun in the field-of-view of about 1.3~$R_\odot$, the spatial resolution of 1.5$\arcsec$ (0.6$\arcsec$~pixel$^{-1}$) and the time cadence of 12~s, in ten extreme ultraviolet (EUV) and UV channels. The present study utilizes 171~\AA\ (\ion{Fe}{9}, corresponding to the temperature $T$ of about 0.7~MK), 94~\AA\ (\ion{Fe}{18}, $T\approx6.3$~MK), 131~\AA~ (\ion{Fe}{8}, \ion{Fe}{21}, \ion{Fe}{23}, i.e., $T\approx 0.4$, 10 and 16~MK, respectively), 304~\AA~(\ion{He}{2}, $T\approx0.05$~MK), and 1,600~\AA\ (\ion{C}{4} + continuum, $T\approx 0.1$~MK and 5000 K) images. We also used Heliospheric and Magnetic
Imager (HMI) magnetograms \citep{schou2012} to investigate the magnetic configuration of the active region. We used RHESSI \citep{lin2002} hard X-ray (HXR) and Nobeyama Radioheliograph (NoRH; \citealt{nakajima1994}) and Polarimeter (NoRP; \citealt{nakajima1985}) observations to determine the HXR and microwave source location of the observed QPP.  

The active region (AR) NOAA 11302 was located near the eastern limb (N12E56) on 23 September 2011, with a  $\beta\gamma$ magnetic configuration. The QPP reported here, was observed during an M1.9 flare. The flare was triggered at the edge of the AR, where the fan-shaped loops were observed within the pre-existing active region. The flare started at $\sim$23:48~UT, peaked at $\sim$23:56~UT, and ended at $\sim$00:04~UT.

\begin{figure*}
\centering{
\includegraphics[width=8.5cm]{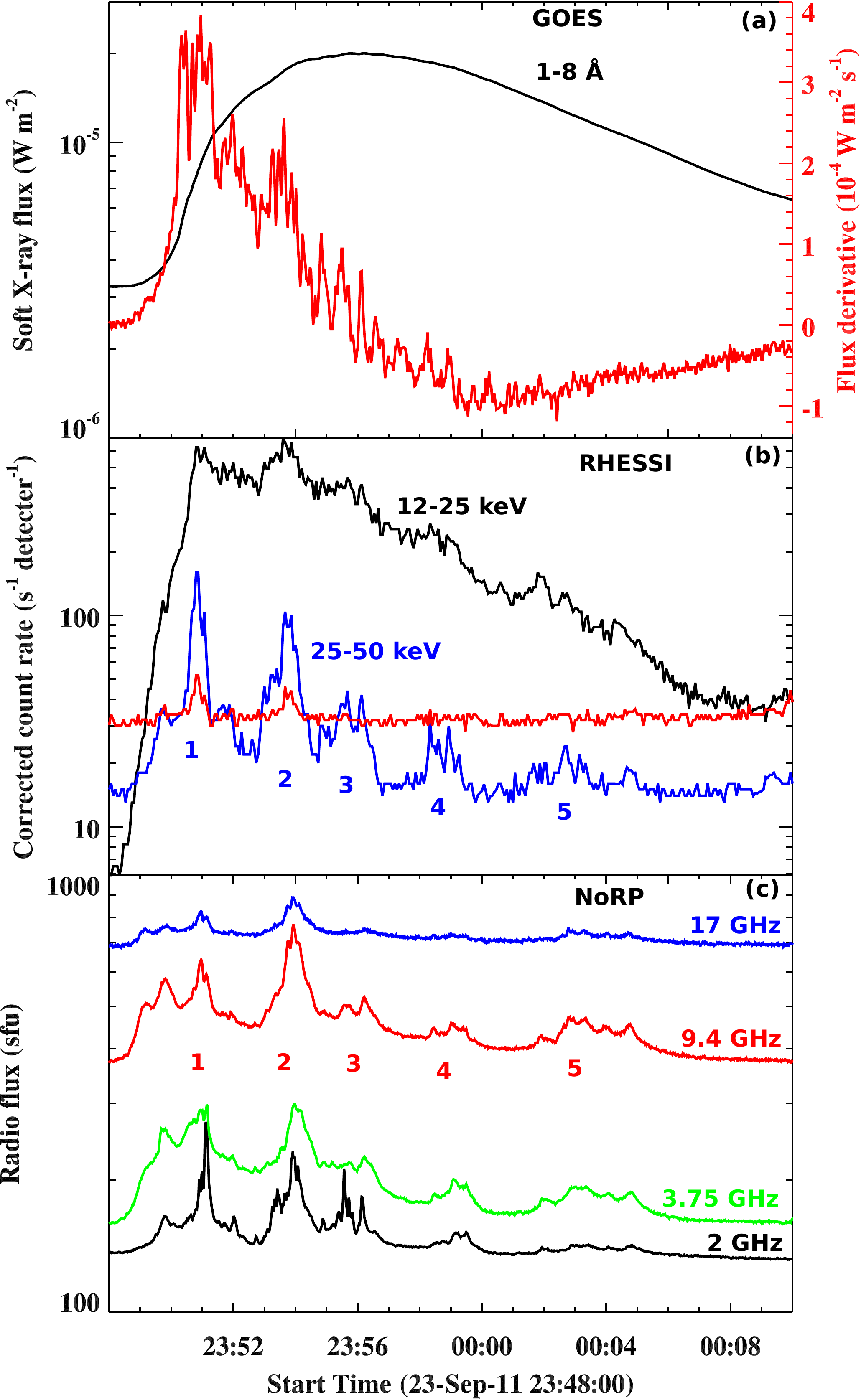}
\includegraphics[width=7.9cm]{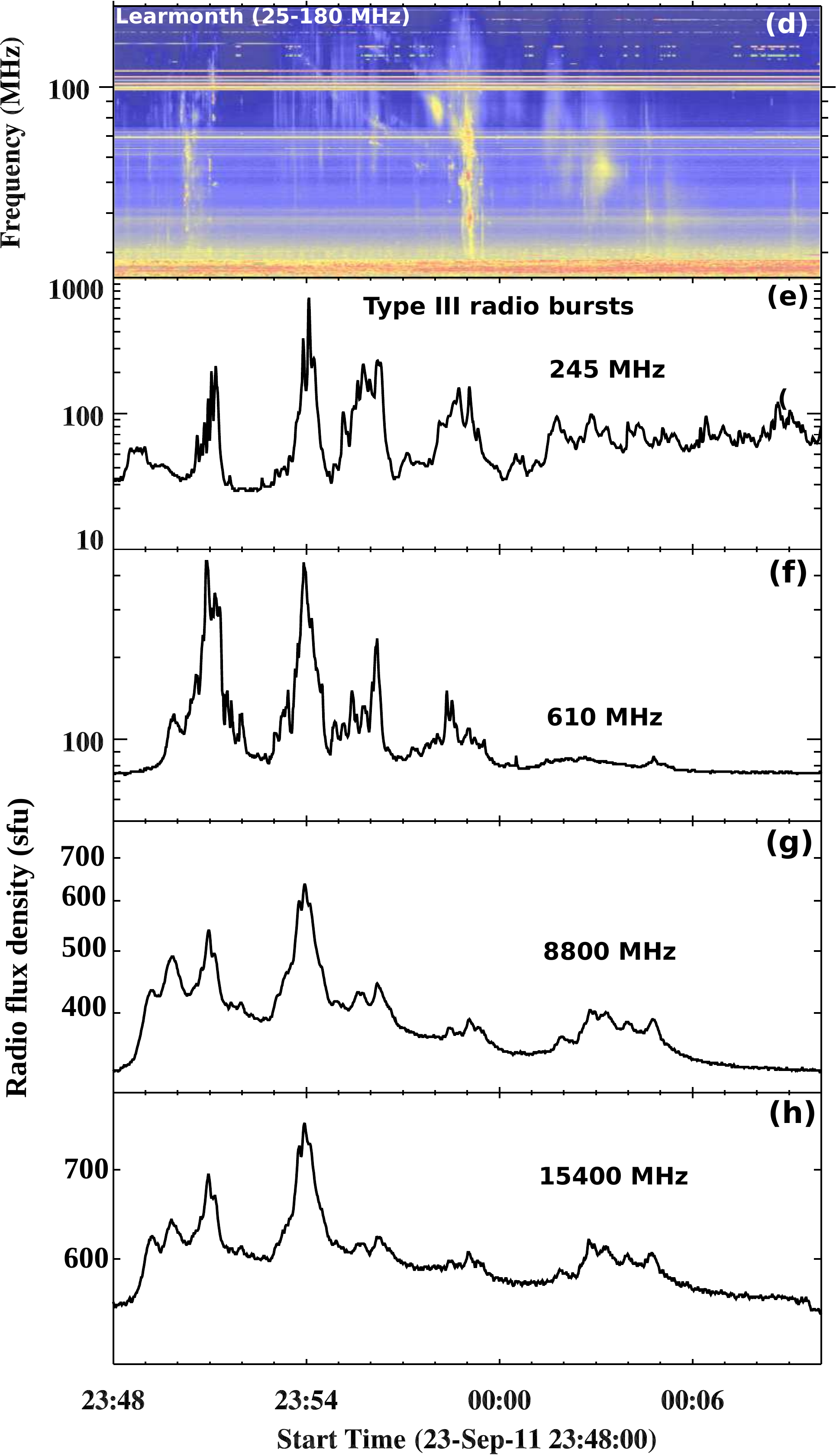}
}
\caption{{\small (a) GOES soft X-ray flux profile in the 1--8 \AA~ channel and its time derivative (red). (b) RHESSI hard X-ray flux profiles in 12--25, 25--50, and 50--100 keV channels. (c) NoRP flux profiles in 2, 3.75, 9.4, and 17~GHz frequency bands. (d) Learmonth dynamic radio spectrum in 25--18~MHz frequency, showing bunches of type III radio bursts. (e-h) RSTN radio flux density profiles in 245, 610, 8,800, 15,400~MHz frequencies from the Learmonth solar observatory.}}
\label{flux}
\end{figure*}

\begin{figure*}
\centering{
\includegraphics[width=7cm]{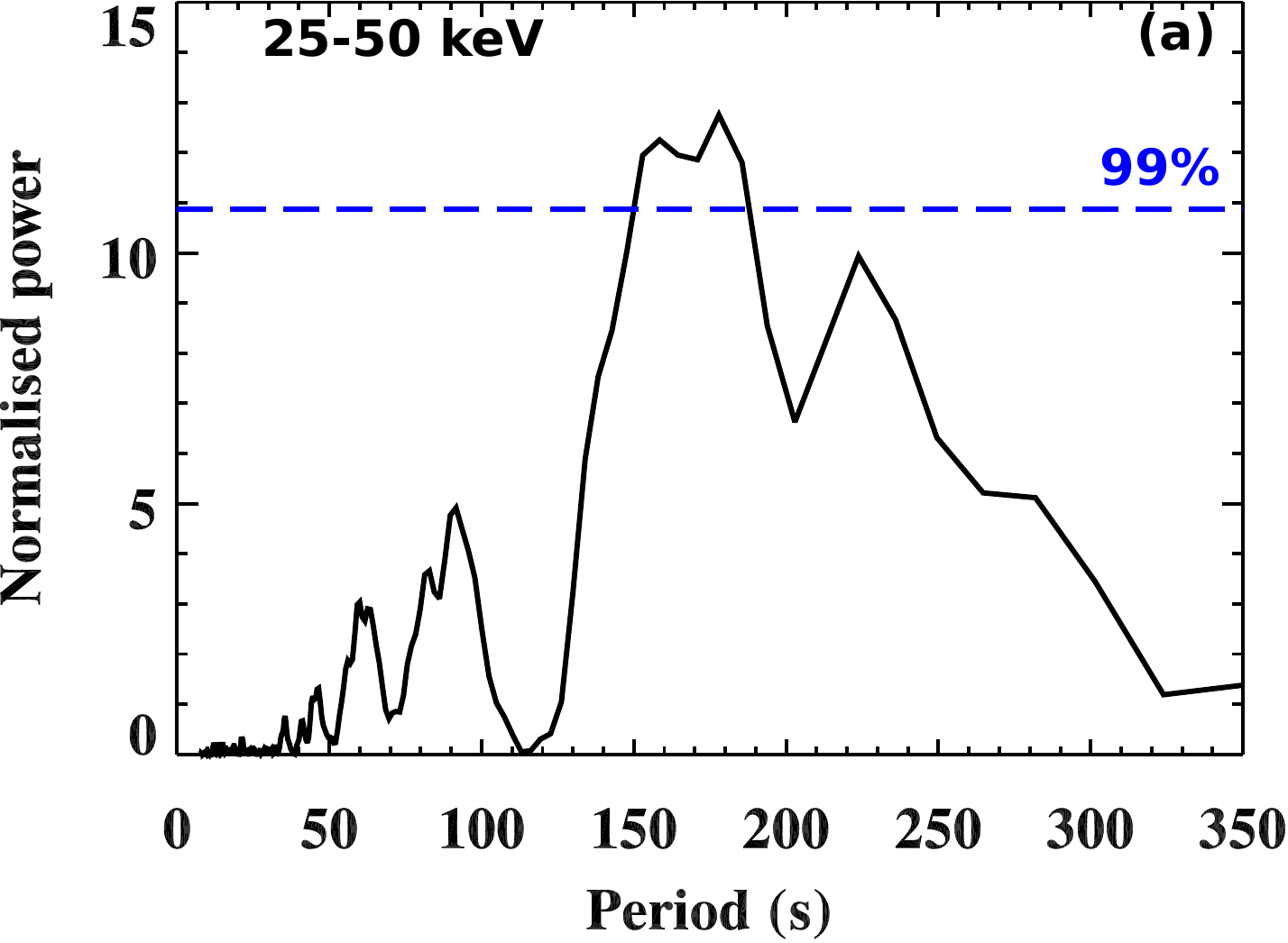}
\includegraphics[width=7cm]{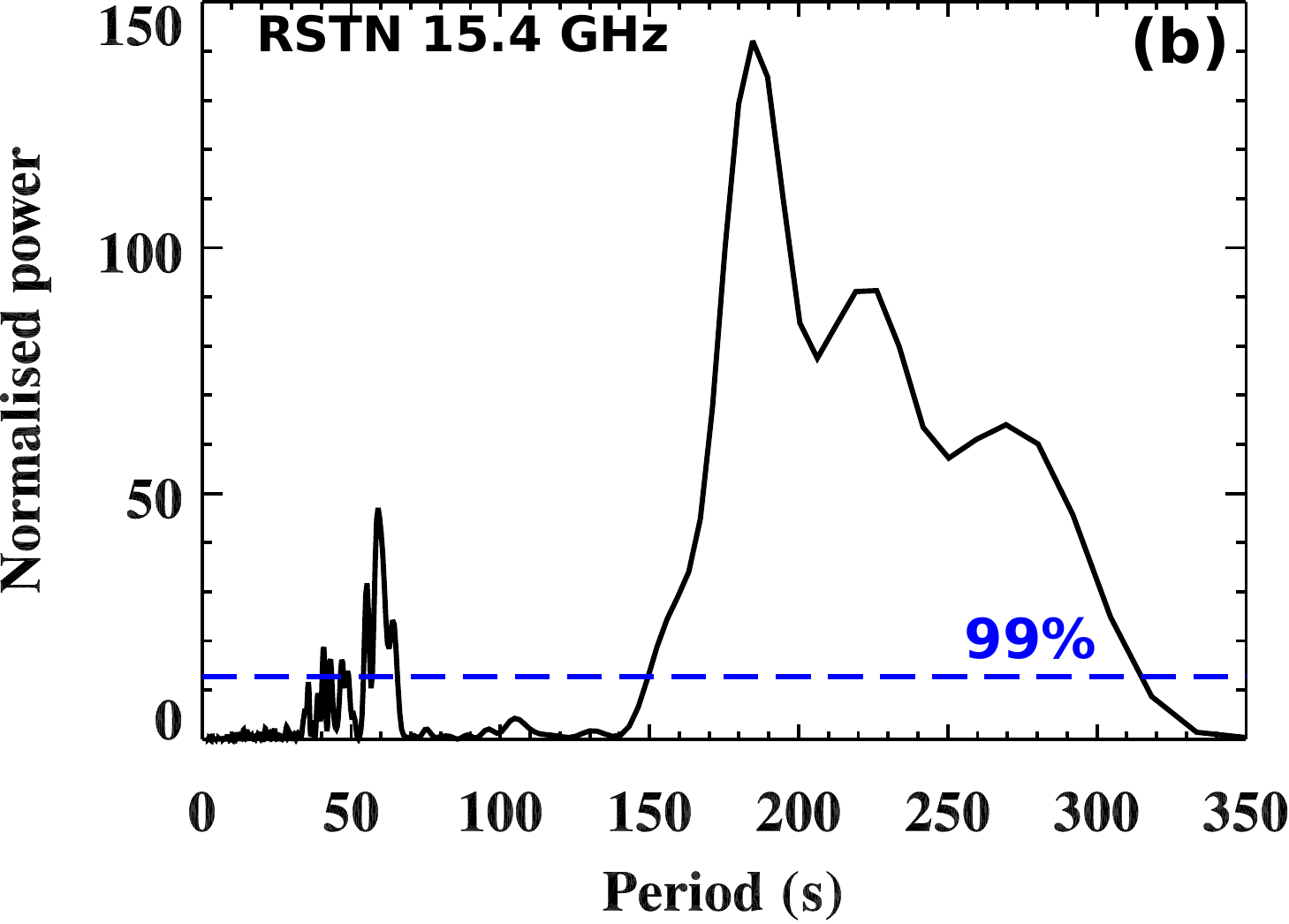}

\includegraphics[width=7cm]{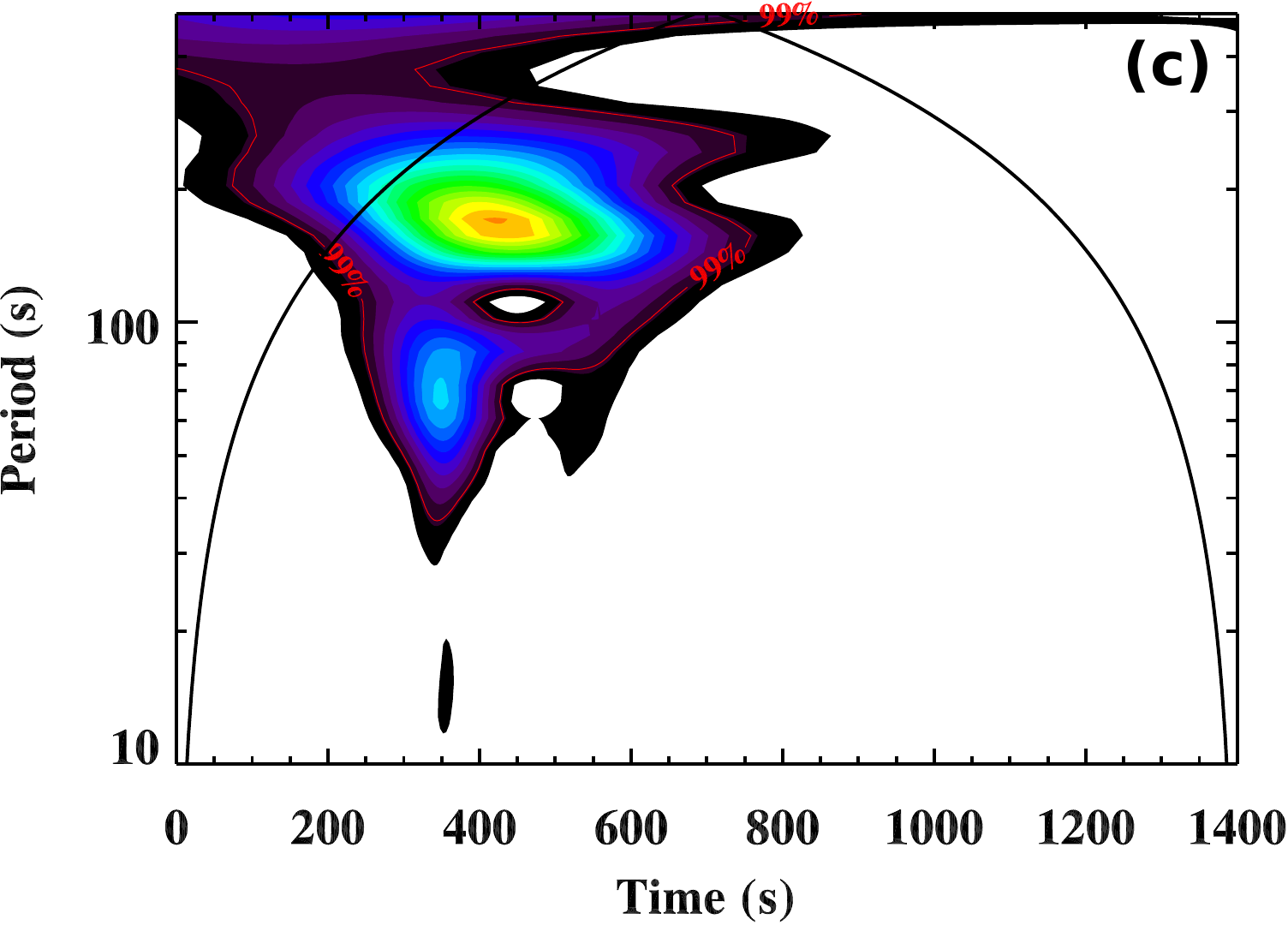}
\includegraphics[width=7cm]{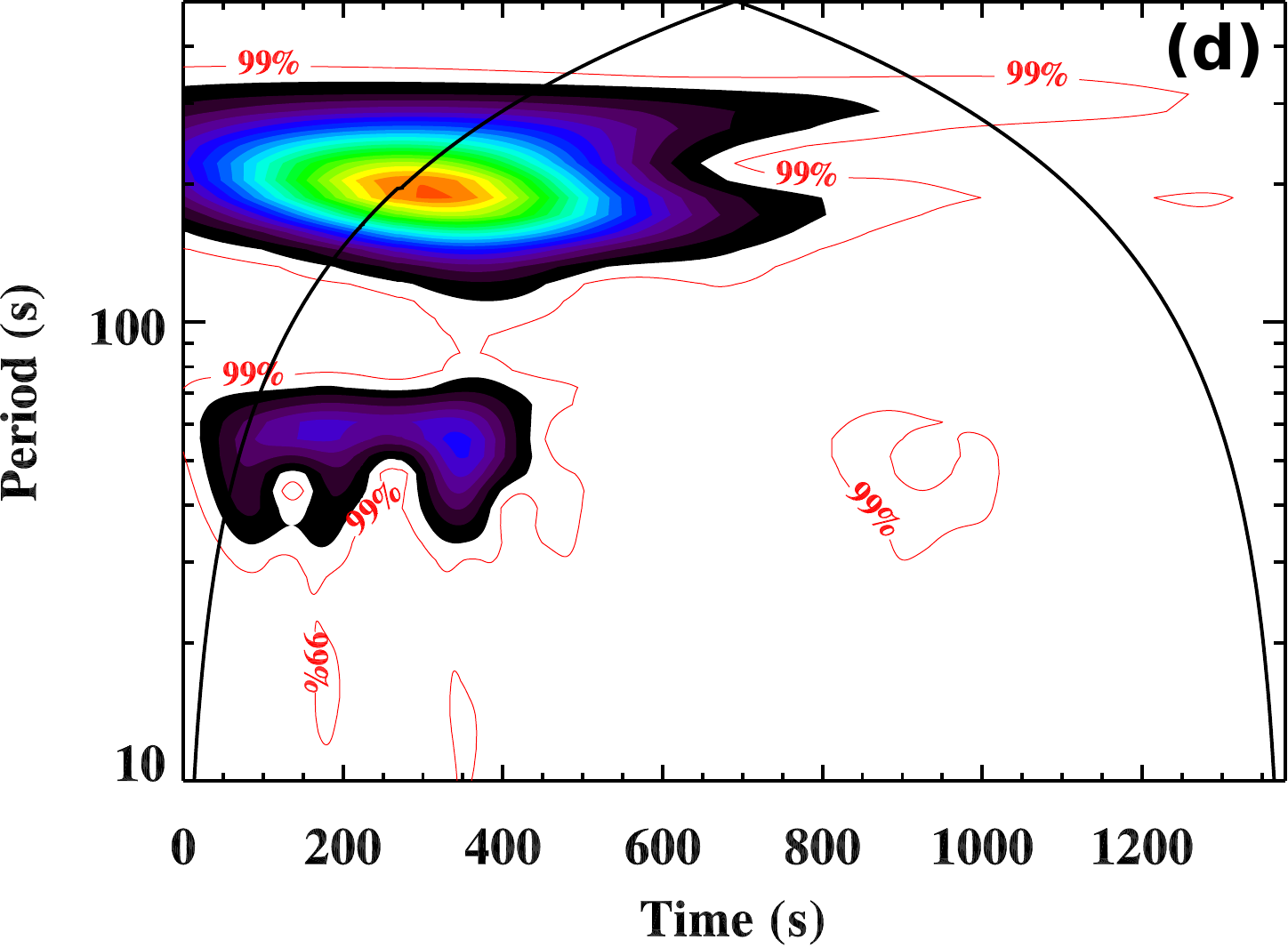}
}
\caption{{\small (a-b) Lomb--Scargle periodograms of RHESSI 25--50~keV and RSTN 15.4~GHz signals. (c-d) Wavelet power spectra for RHESSI 25--50~keV energy channel (left) and RSTN 15.4~GHz (right) signals. The start time is 23:48~UT.}}
\label{period}
\end{figure*}

\begin{figure}
\centering{
\includegraphics[width=8.5cm]{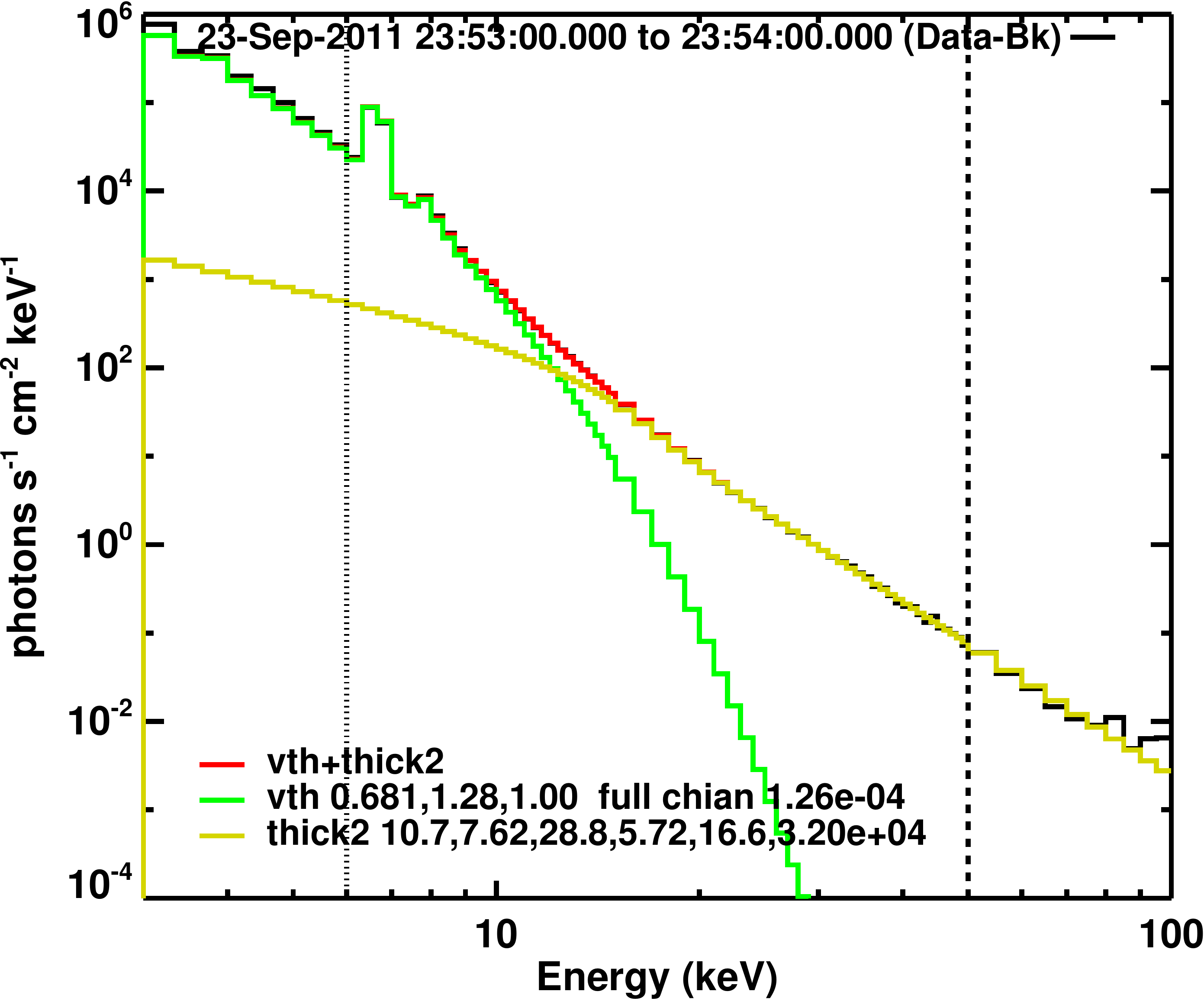}
}
\caption{{\small RHESSI energy spectrum fitted with the isothermal (green) and thick-target bremsstrahlung (yellow) models for the interval 23:53--23:54~UT.}}
\label{spectrum}
\end{figure}

\begin{figure*}
\centering{
\includegraphics[width=7cm]{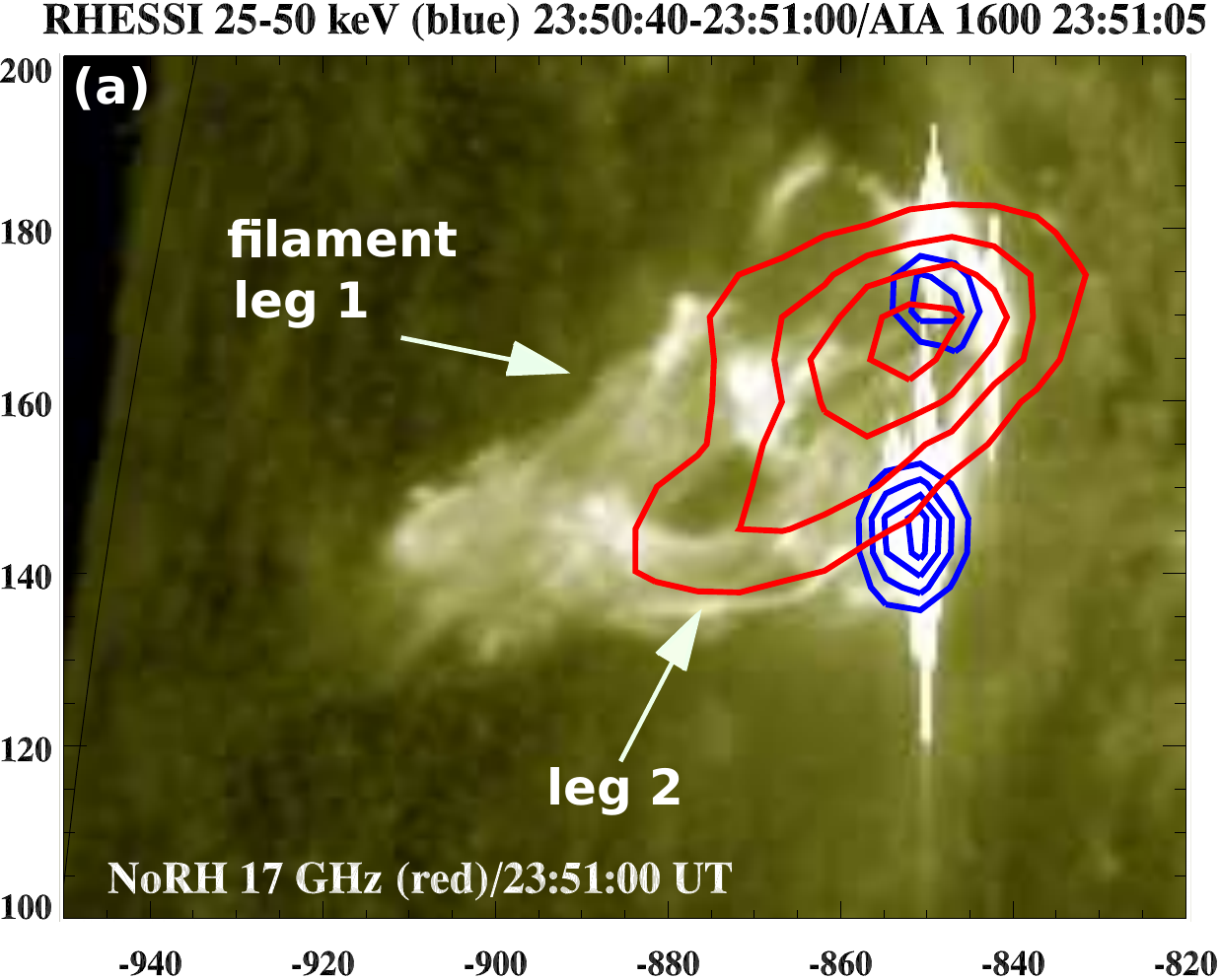}
\includegraphics[width=7cm]{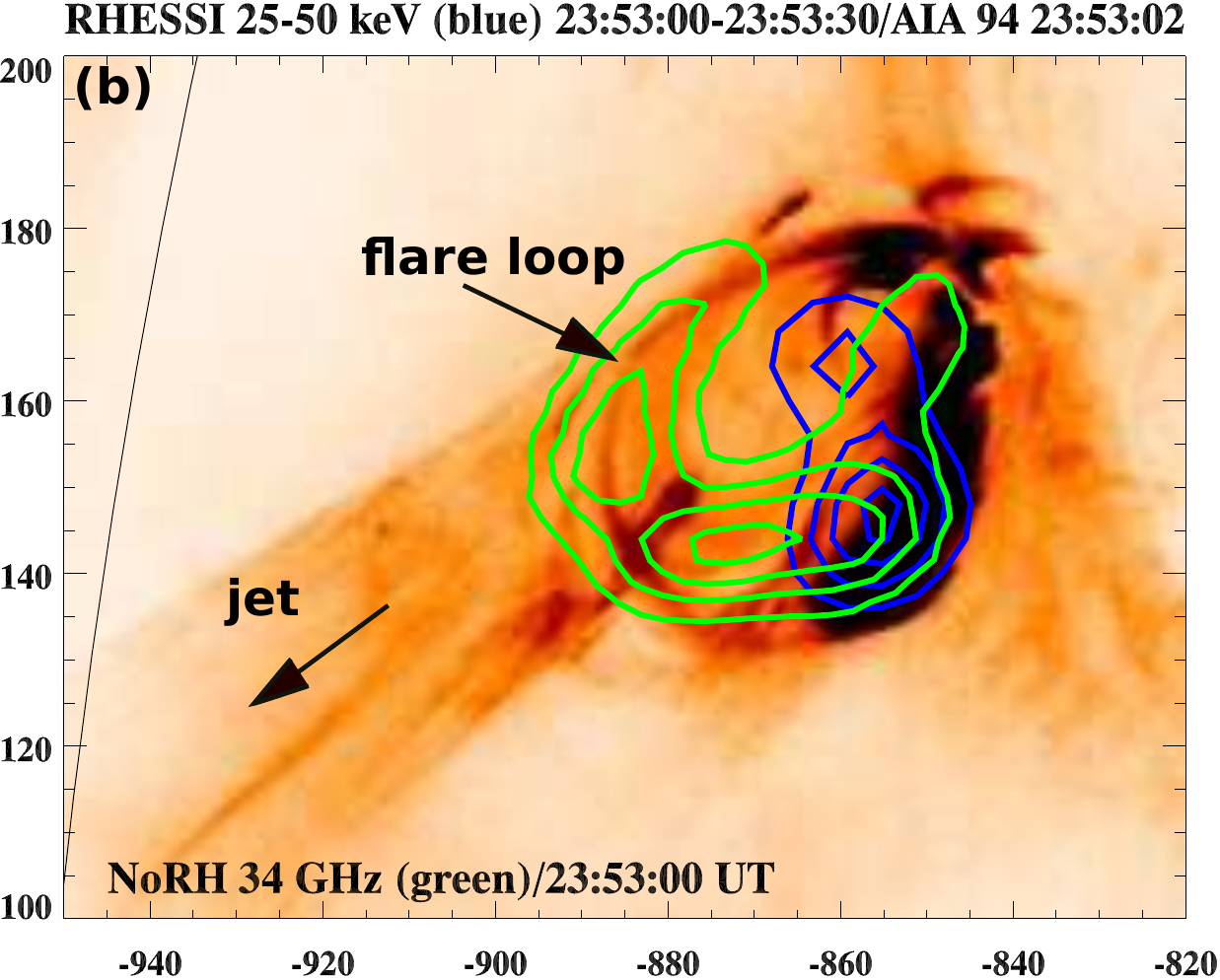}

\includegraphics[width=7cm]{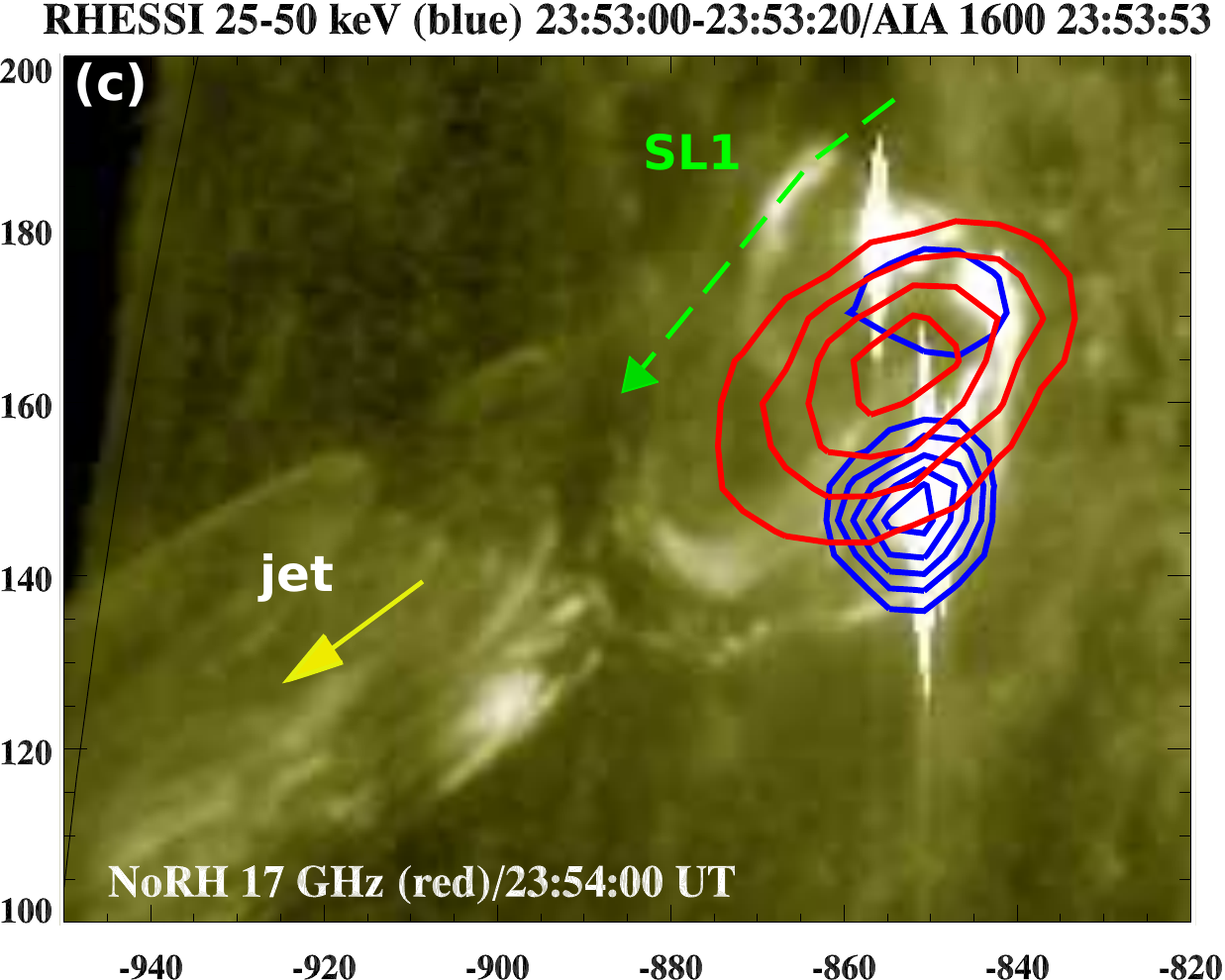}
\includegraphics[width=7cm]{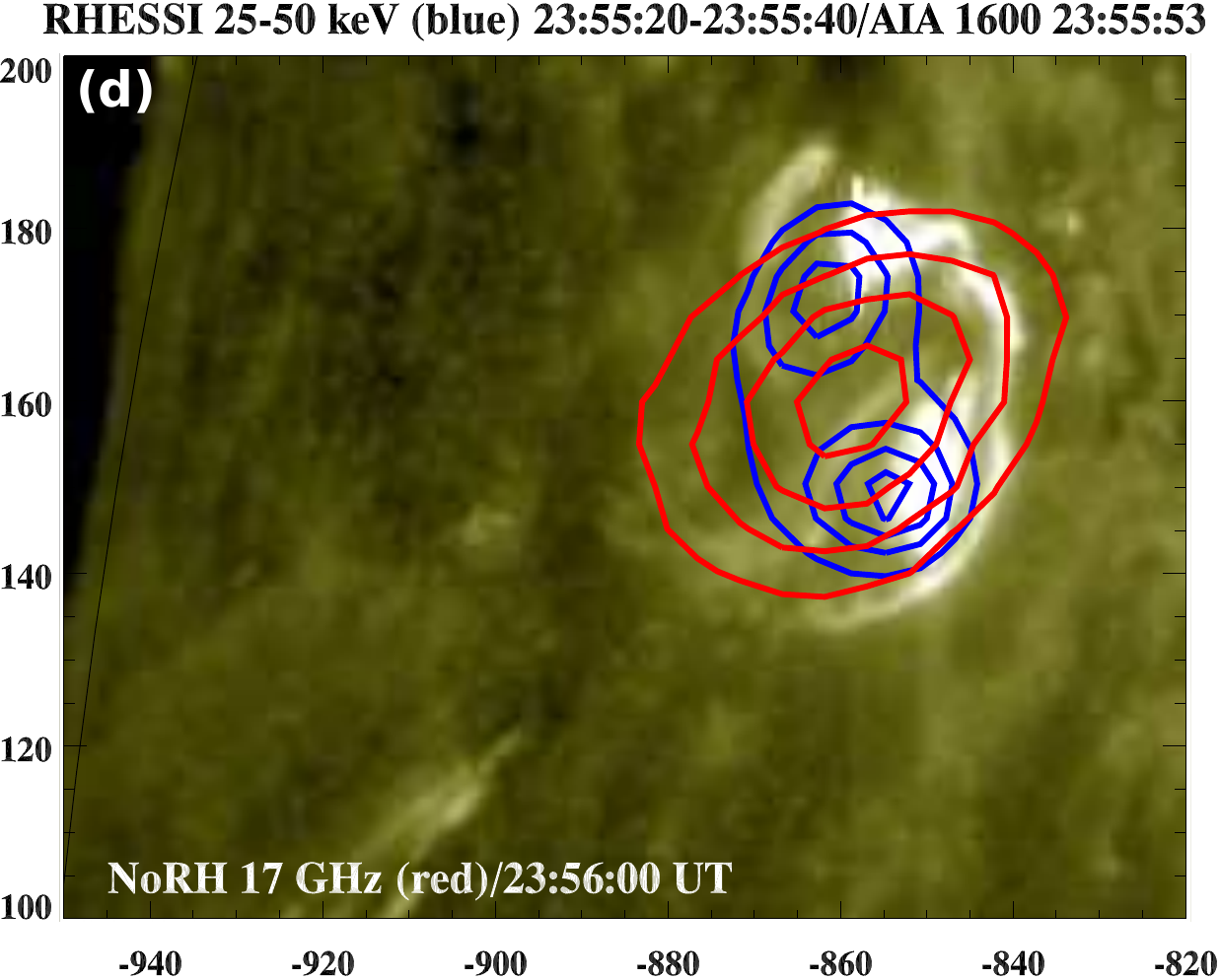}

\includegraphics[width=7cm]{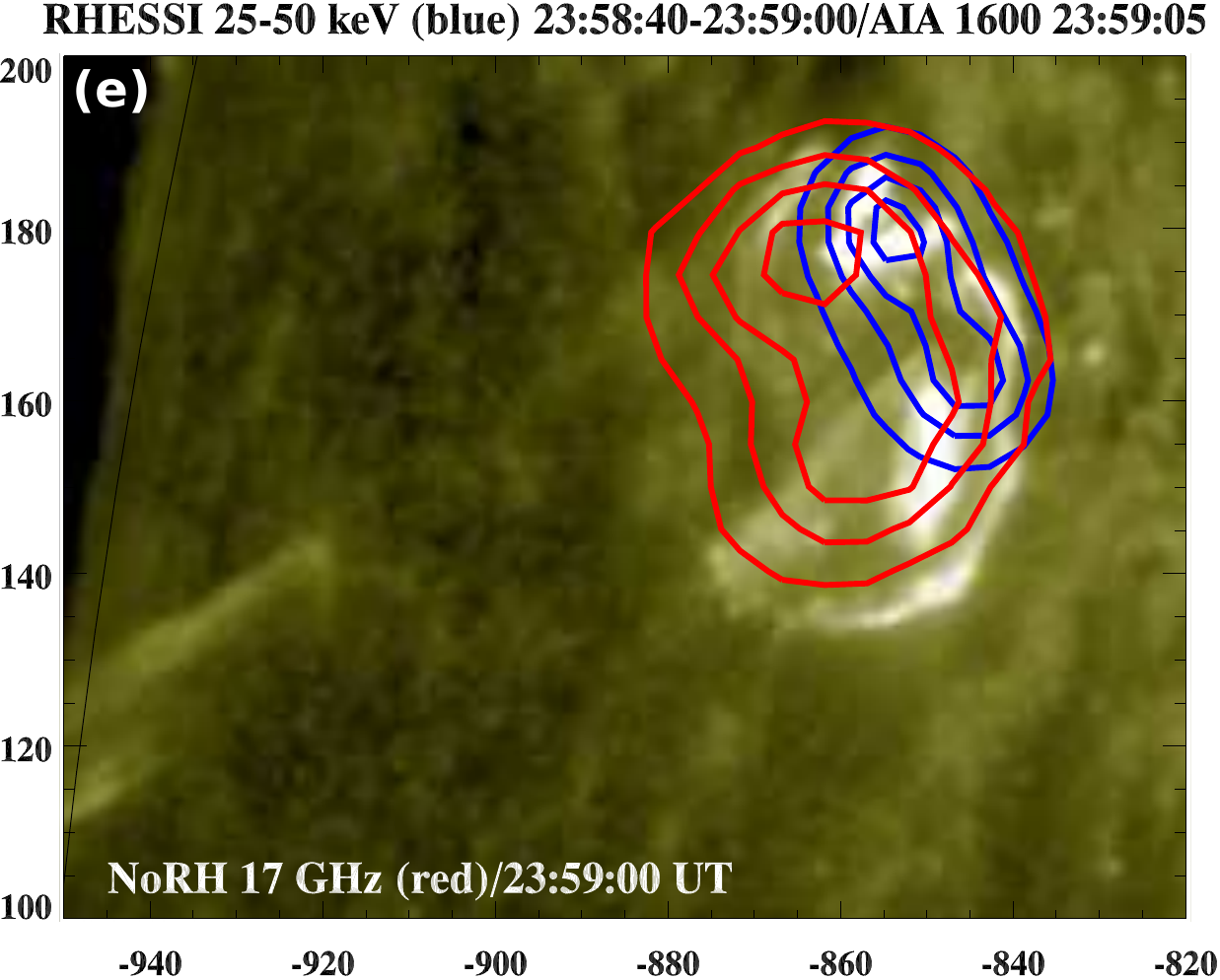}
\includegraphics[width=7cm]{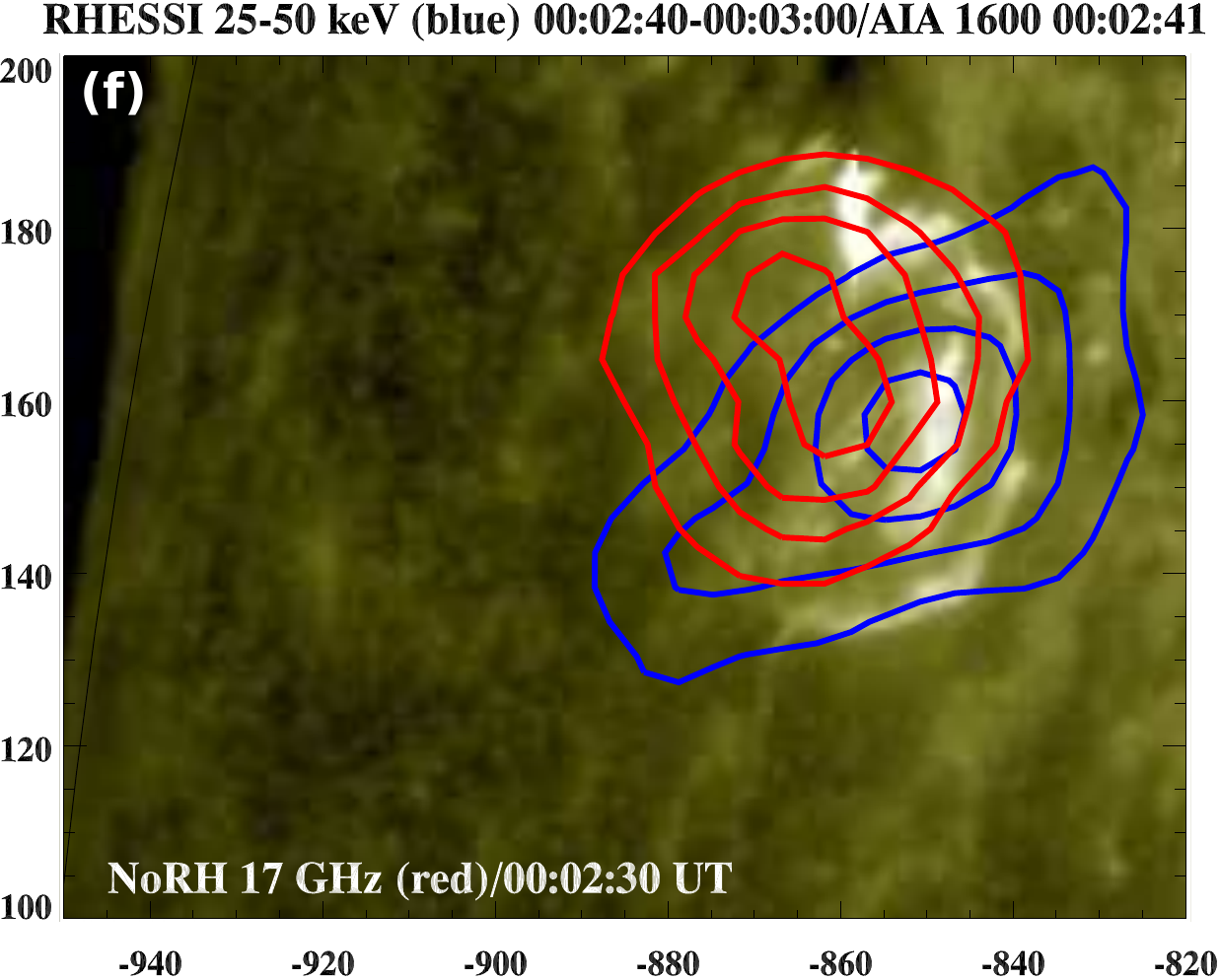}
}
\caption{{\small RHESSI hard X-ray (25--50~keV, blue) and NoRH (R+L) 17~GHz sources shown by the contours overlaid on the AIA 1,600 \AA~ images. Panel (b) shows RHESSI 25--50~keV (blue) and NoRH 34~GHz (green) contours over the AIA 94 \AA~ negative image. The time of each image matches the peak time of the 3-min oscillation in the hard X-ray 25--50~keV and radio 17~GHz. The contour levels are 30\%, 50\%, 70\%, 90\% of the peak flux. We included an extra contour level (i.e., 10\% of the peak flux in panel (c)) to show the opposite foot-point of the loop. The axes are labelled in arcsecs.}}
\label{source}
\end{figure*}


\subsection{Quasi-periodic pulsation}

Figure \ref{flux}(a) displays the GOES soft X-ray flux profile (3-s cadence) in 1--8 \AA~ channel and its derivative (red curve) estimated by using the 3-point Lagrangian interpolation method. Interestingly, the time derivative of the flux profile reveals a clear QPP. Fig.~\ref{flux}(b) shows RHESSI hard X-ray count rate time profiles (4-s cadence) in the 12--25 keV (black), 25--50 keV (blue), and 50--100 (red) keV channels. The decaying oscillation is especially very clear in the 25--50 keV channel. Five distinct bursts of the energy release are marked by 1 to 5. The 50--100 keV flux profile shows only first two bursts of the energy release, while the other three bursts are not detected in this energy range. To investigate the QPP in radio channels, we plotted Nobeyama Radio Polarimeter (NoRP, 1-s cadence) flux profiles in 2, 3.75, 9.4, and 17~GHz channels (Fig.~\ref{flux}(c)). We can see a clear QPP cotemporal with the RHESSI HXR bursts.

Fig.~\ref{flux}(d) displays the dynamic radio spectrum in 25--180 MHz frequency observed at the Learmonth radio observatory, Australia. We can see a series of quasi-periodic type III radio bursts during the flare. We also used  Radio  Solar Telescope  Network (RSTN) 1-s cadence data from the Learmonth observatory to see the radio flux density profiles in 245, 610, 8,800, and 15,400~MHz frequency bands (Fig.~\ref{flux}(d-h)). QPP is clear in all the light curves not only in the high frequency radio bursts (2, 9.4, 17~GHz) but also in the low frequency (245, 610~MHz) channels.


Nonthermal electrons responsible for the HXR and microwave emissions may be accelerated by the same process during the flare, e.g. magnetic reconnection \citep{kundu1961,kundu1965}, but have different energies \citep{kai1986}. The foot-point HXR emission is generated during the precipitation of the nonthermal electrons with the energies of tens of keV, whereas the microwave emission is produced by trapped nonthermal electrons of several hundreds of keV \citep{kundu1994,white2003,asai2013}. Spatially, the microwave emission comes from the body of the flaring loop, while the HXR emission comes from its footpoints and loop-top. Therefore, microwave and HXR emissions are generated by the nonthermal electrons accelerated in the reconnection site and precipitated/transported downward into the chromosphere. In contrast, the 25--180 and 245~MHz channels show quasi-periodic type III radio bursts excited by the nonthermal electrons propagating upward (along the open field lines) from the reconnection site to the interplanetary medium. The emission at the decimetric frequency (e.g. 610 MHz) is generally associated with the acceleration region \citep{asc1997,asc2004}. Type III radio bursts above 1 GHz are generated by the electron beams propagating downward from the acceleration site \citep{bastian1998}. Since we observed $\sim$3-min periodicity in type III radio bursts (below 300 MHz), microwave bursts (above 1 GHz), and HXR bursts (25-50 keV) simultaneously, therefore,  it indicates the bi-directional acceleration/injection of electrons \citep{asc1997} from the acceleration region. Thus, this event suggests the quasi-periodic acceleration of electrons, that then propagate from the reconnection site upwards and downwards, bidirectionally. 

Although the oscillatory pattern is clear in all the original light curves, we used the Lomb--Scargle periodogram method \citep[e.g.][]{scargle1982,horne1986} for the determination of the oscillation period and its confidence level. Figure~\ref{period}(a,b) shows the results of the periodogram analysis (i.e., the power spectra) for the 25--50~keV RHESSI hard X-ray signal (4-s cadence) and RSTN 15.4~GHz radio signal (1-s cadence). The confidence level of 99\% is marked by the horizontal dashed line. A period of approximately 3~min (above 99\% confidence level) is revealed in both the RHESSI and microwave signals. 
We would like to stress that the confidence levels shown in the power spectra correspond to the highest spectral peaks only, and cannot be used for the determination of the significance of the secondary peaks.
We also applied Morlet wavelet analysis \citep{torrence1998} on the RHESSI and NoRH light curves. The resultant wavelet power spectra are shown in Fig.~\ref{period}(c,d). Again the 3-min periodicity is seen to be significant in both the wavelet spectra above the 99\% confidence level, which is consistent with the periodogram analysis. 

We generated the X-ray spectrum obtained with  RHESSI during 23:53-23:54~UT using the OSPEX package in SSWIDL (Figure~\ref{spectrum}).
 The background-subtracted photon flux is fitted with thermal (green) and nonthermal (yellow) components. We used an isothermal component (V$_{th}$) for the optically thin thermal bremsstrahlung radiation and thick target bremsstrahlung version-2 model (thick2) for the nonthermal photon flux due to the interaction of accelerated electrons with the thick target plasma. We chose 6--50~keV energy range indicated by vertical dotted lines, for the fitting purpose.  From the best-fitting parameters, we derived the temperature $T$ and emission measure (EM) of the hot plasma to be 14.8~MK and 0.68$\times$10$^{49}$~cm$^{-3}$, respectively. The spectral index for the thick-target bremsstrahlung is 5.7. The nonthermal component dominates  over the thermal component above $\sim$13~keV. Therefore, it is clear that the observed QPP is basically related to the periodic variation of the nonthermal electron flux.

\subsection{Spatial location of the hard X-ray and radio sources}
To investigate the particle precipitation/transport sites during the flare, we used  hard X-ray 25--50~keV and NoRH 17/34~GHz images. We chose the PIXON algorithm \citep{metcalf1996} for the RHESSI image reconstruction. The pixon method is considered to be the most accurate algorithm \citep{hurford2002}. The integration time for each image was 20~s. We utilised NoRH 5-s cadence intensity images (R+L) at 17 and 34~GHz. Figure~\ref{source} displays the hard X-ray 25--50~keV (blue) and NoRH 17~GHz (red) contours overlaid on the AIA 1,600 \AA~ images at $\sim$23:51~UT. These images are used at the peak time of each of the bursts observed in the hard X-ray 25--50~keV and microwave channels. Figure~\ref{source}(a) shows a small filament rising at the flare site. We can easily identify the two legs of the filament (marked by N and S in the figure). Generally, the filaments are observed in the chromospheric images (e.g., H$\alpha$ and AIA 304 \AA. However, if the kink unstable filament is heated during magnetic reconnection, it is often observed in the AIA 1600 \AA~ channels (e.g., \citealt{kumar2012a} and \citealt{kumar2014}). The locations of both the 25--50~keV sources and 17~GHz source are almost at the quasi-circular ribbon. However, their centroids constructed at the 90\% level of the peak intensity are not cospatial.  We note that the 25--50~keV sources (centroid) are located close to the legs of the filament, whereas 17~GHz source is located at the northern leg of the filament. It seems that these sources are at the footpoints of an underlying flare loop. To identify the location of the flare loop, we selected the AIA 94 \AA~ hot channel image at 23:53 UT. In this image (panel b), we showed NoRH 34~GHz contours (yellow) overlaid on AIA 94 \AA~ images at $\sim$23:53 UT. This is done to show the coronal loops associated with the eruption of the small filament. The overlying 34~GHz source is cospatial with the small loop located above the quasi-circular ribbon. This 34~GHz emission may be the evidence of trapped nonthermal electrons in the loop. The rising filament is heated during reconnection with the ambient fields, and then decays into an untwisting jet as seen in Fig.~\ref{source}(b-e). Furthermore, during $\sim$23:53-23:59~UT, we see the footpoint hard X-ray sources. The NoRH 17 GHz sources cover the quasi-circular ribbon (Figure \ref{source}(c-d)) and could be the emission from the trapped electrons in the loop. This suggests that the bursts are most likely caused by the same population of nonthermal electrons with different energies. 

\begin{figure*}
\centering{
\includegraphics[width=5.5cm]{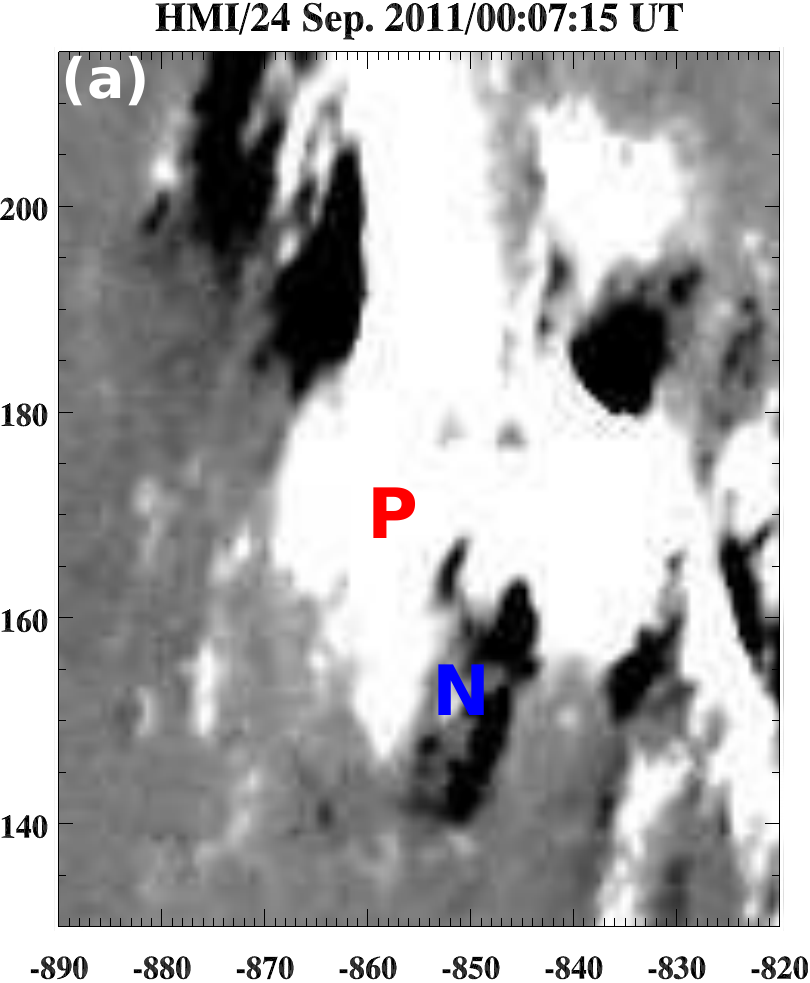}
\includegraphics[width=5.7cm]{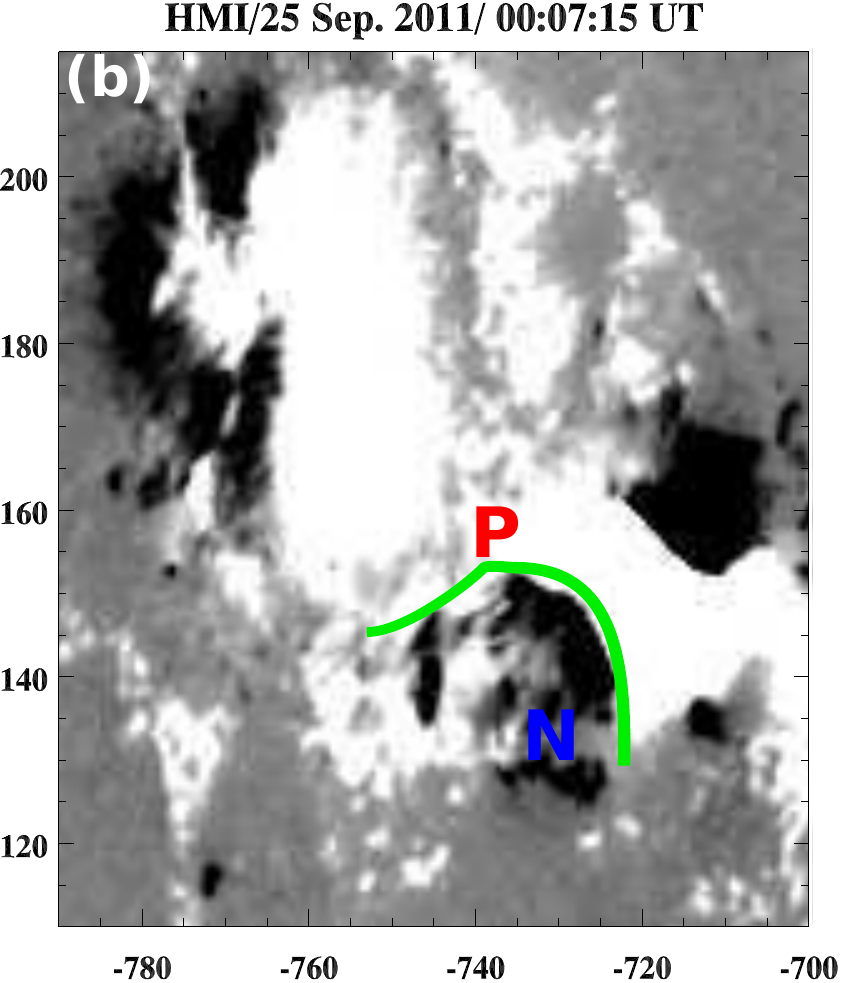}

\includegraphics[width=5.5cm]{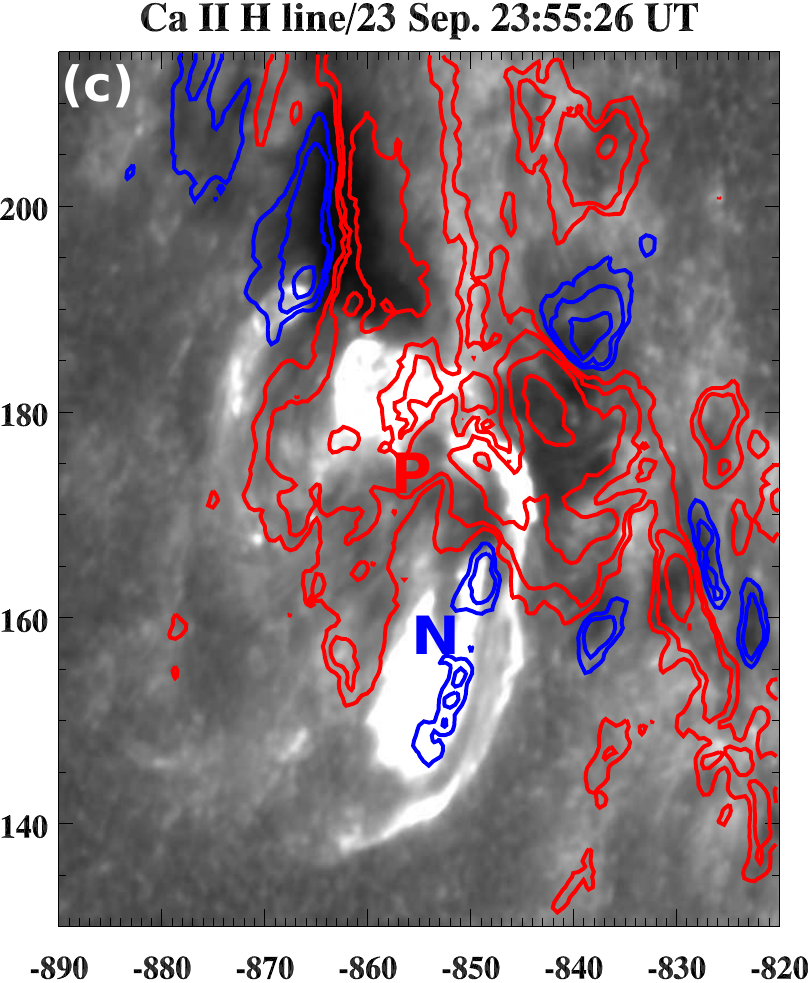}
\includegraphics[width=5.5cm]{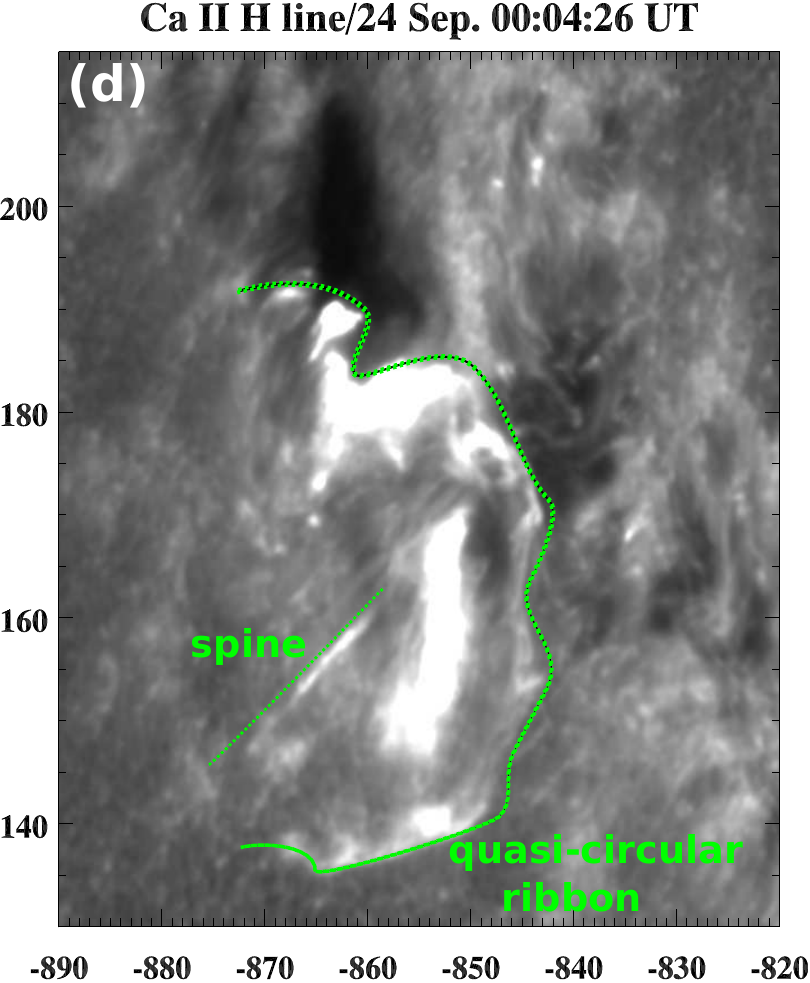}

\includegraphics[width=6.2cm]{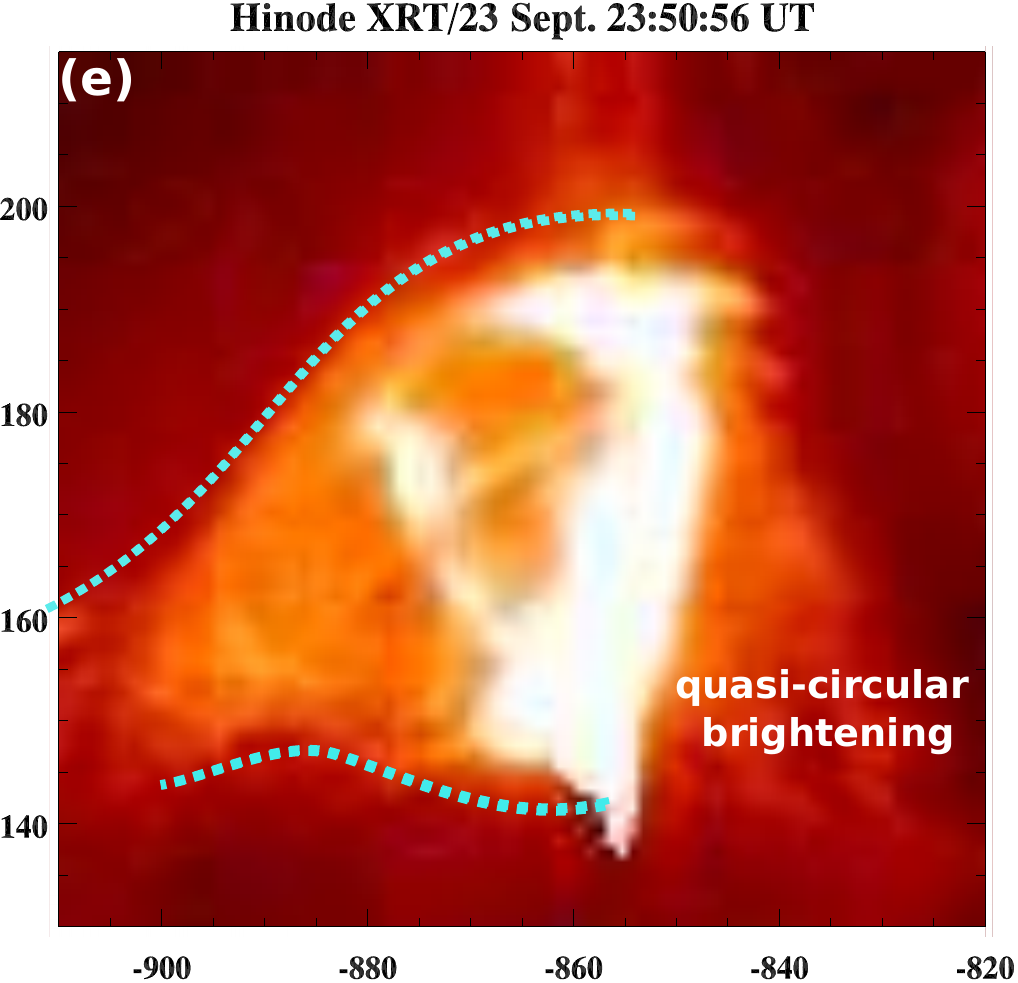}
\includegraphics[width=10.0cm]{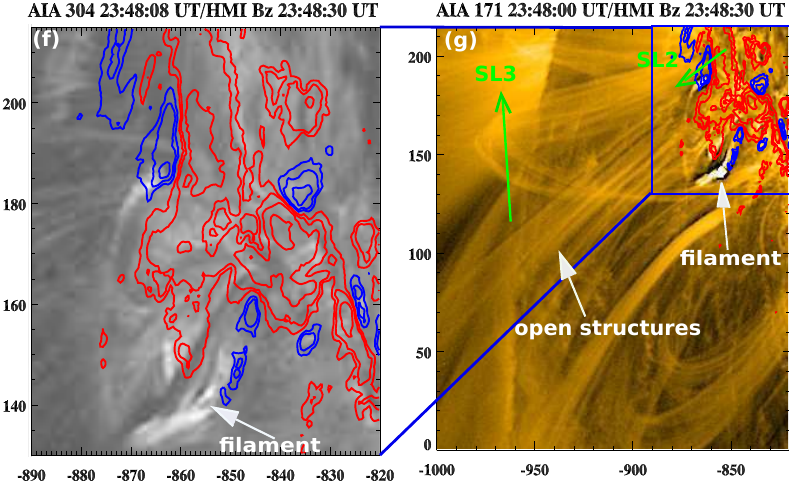}
}
\caption{{\small (a-b) HMI magnetogram of the flaring AR on 24 and 25 September 2011. P and N indicate positive and negative polarity fields. Green line is the PIL. (c-d) Hinode/SOT Ca~II~H line (3968 \AA) images showing the flare ribbon morphology. (e-g) XRT image (Be\_thin filter) during the flare onset showing the magnetic configuration and formation of quasi-circular brightening at the foot-points of the loops. AIA 304 and 171 \AA~ images overlaid by HMI magnetogram of positive (red) and negative (blue) polarities. 
The contour levels are $\pm$200, $\pm$400, $\pm$800, and $\pm$1,600~G.}}
\label{mag}
\end{figure*}


\begin{figure*}
\centering{
\includegraphics[width=17cm]{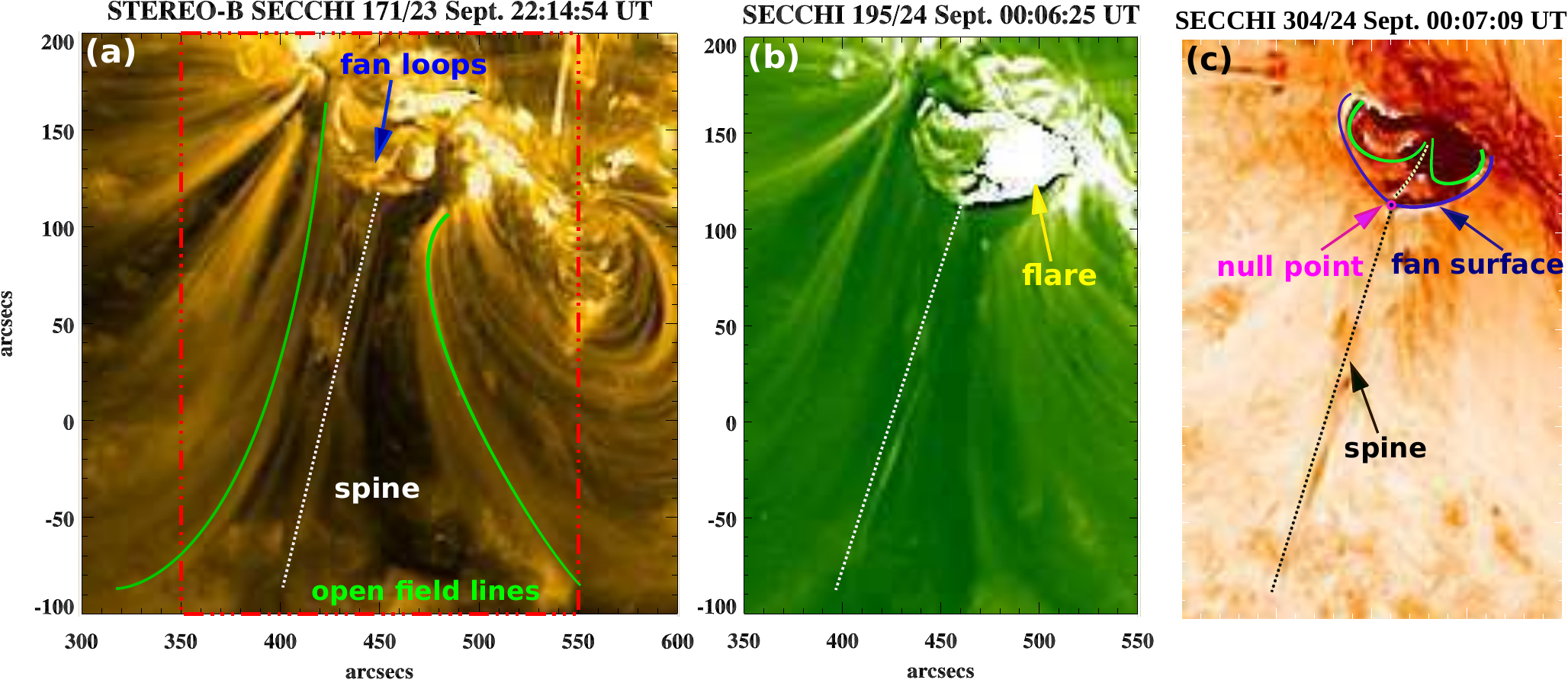}
}
\caption{{\small STEREO-B SECCHI intensity images in 171, 195, and 304 (reverse colour) \AA~ channels. These images display the possible magnetic configuration (i.e., fan-spine topology) of the flare site. The rectangular box in panel (a) indicates the location of the field-of-view in panels (b) and (c).}}
\label{secchi}
\end{figure*}

\begin{figure*}
\centering{
\includegraphics[width=5.3cm]{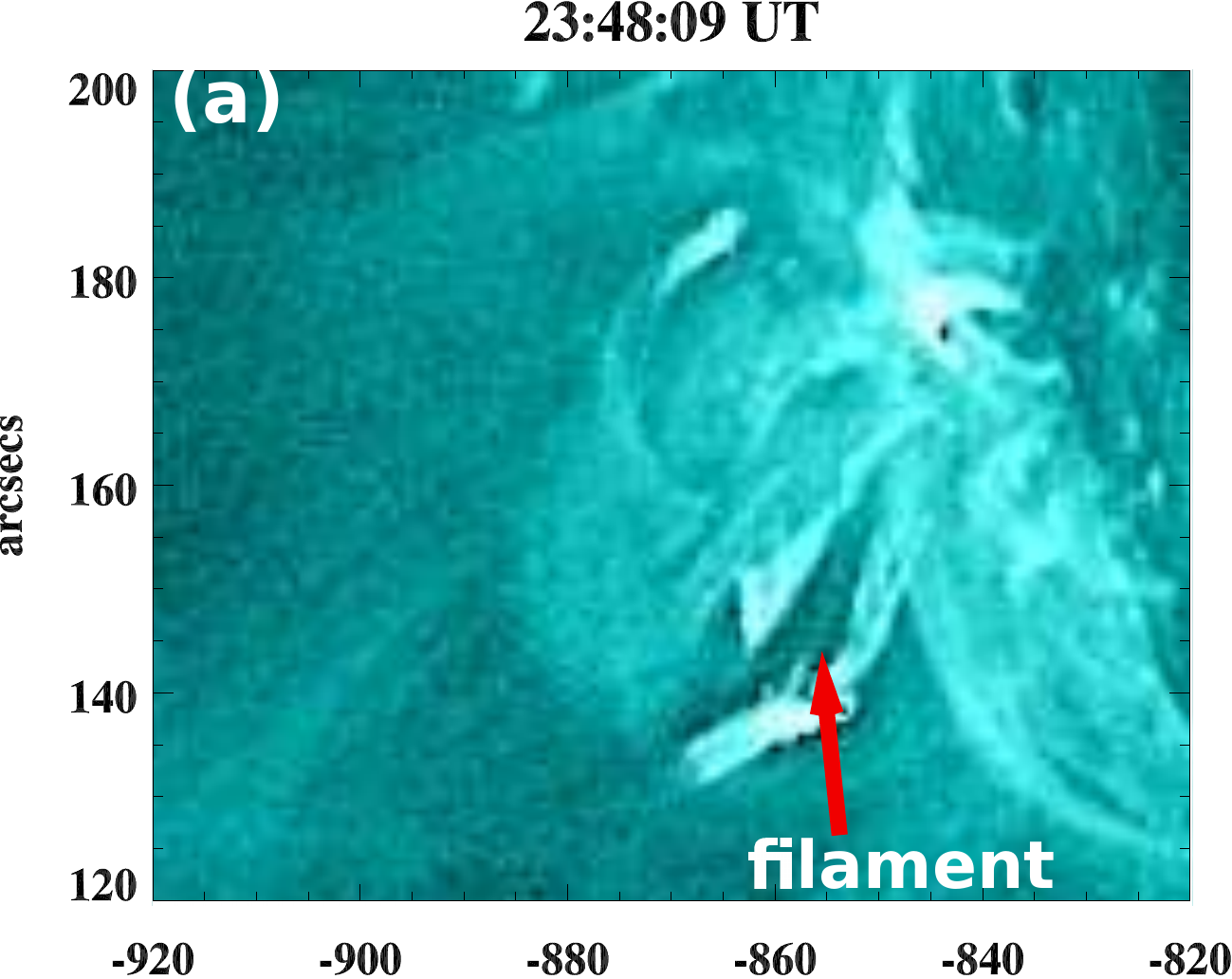}
\includegraphics[width=5.3cm]{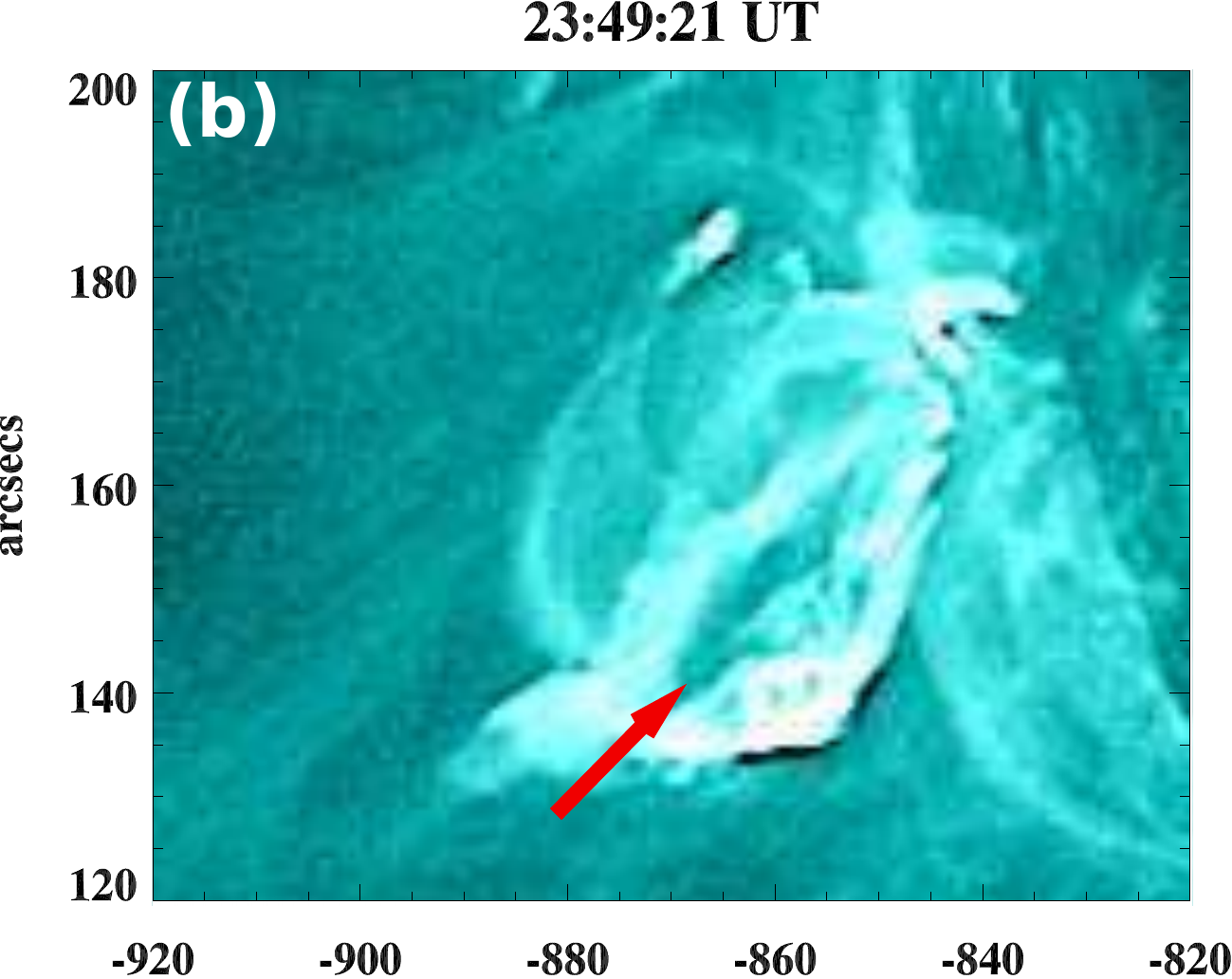}
\includegraphics[width=5.3cm]{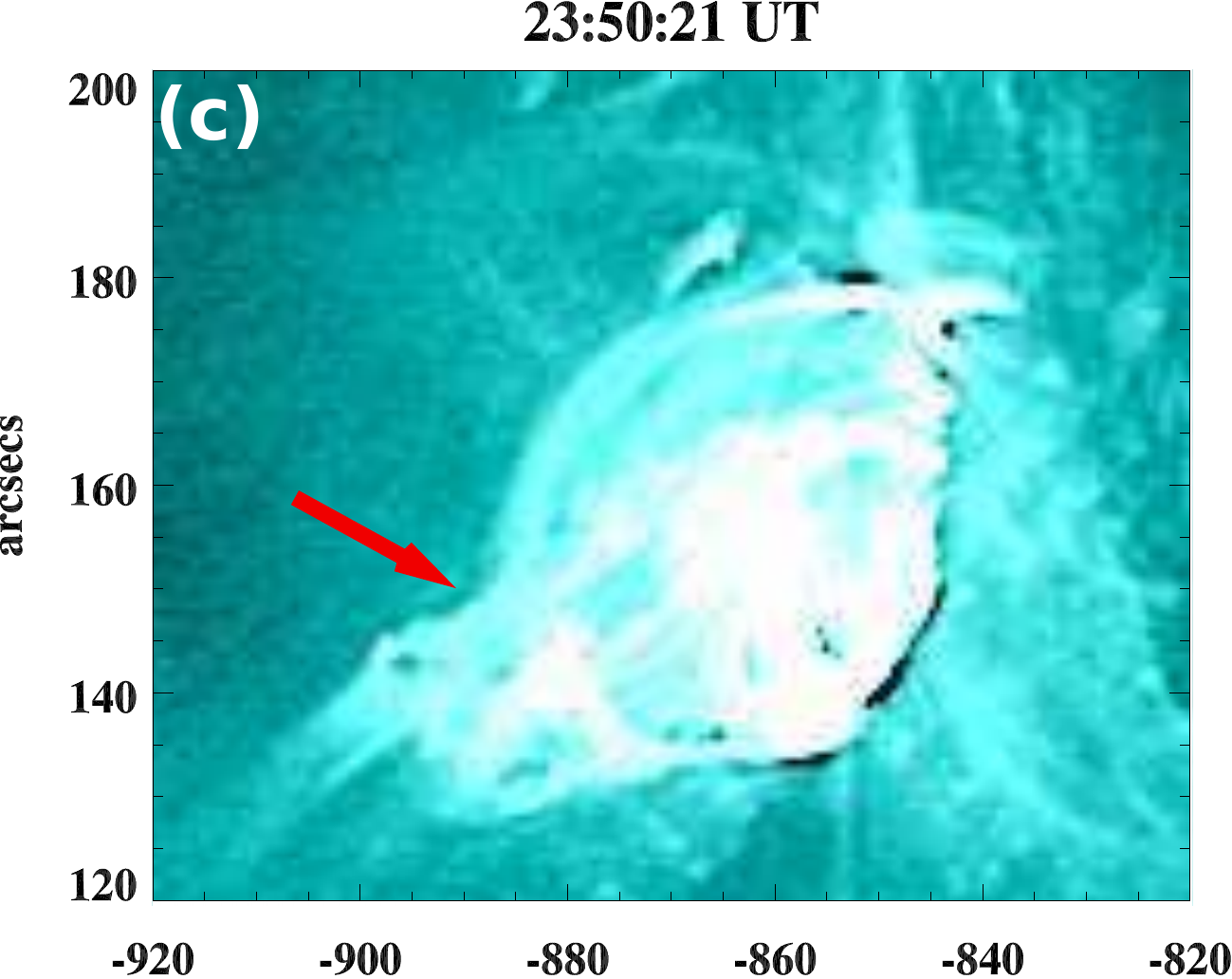}

\includegraphics[width=5.3cm]{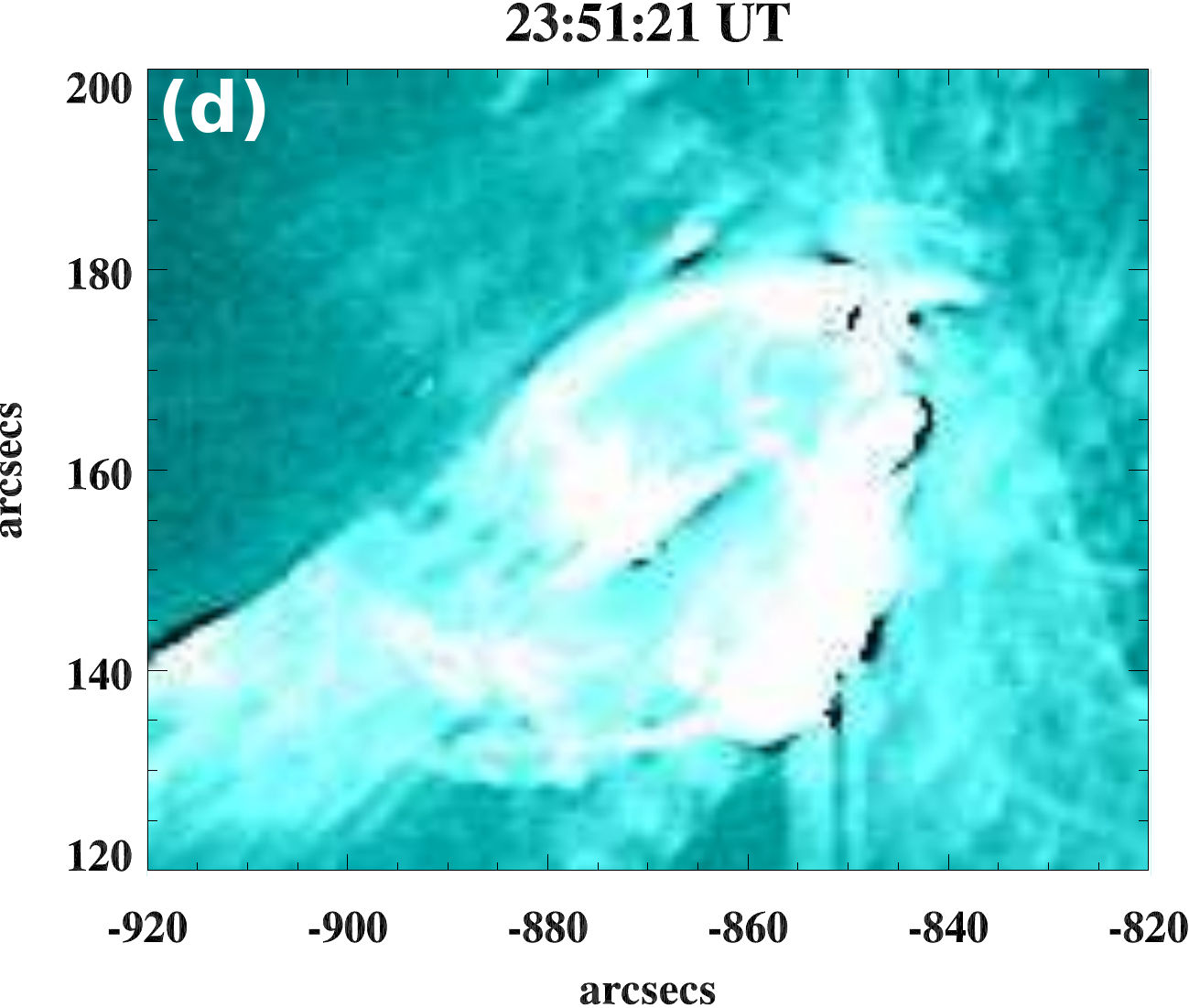}
\includegraphics[width=5.3cm]{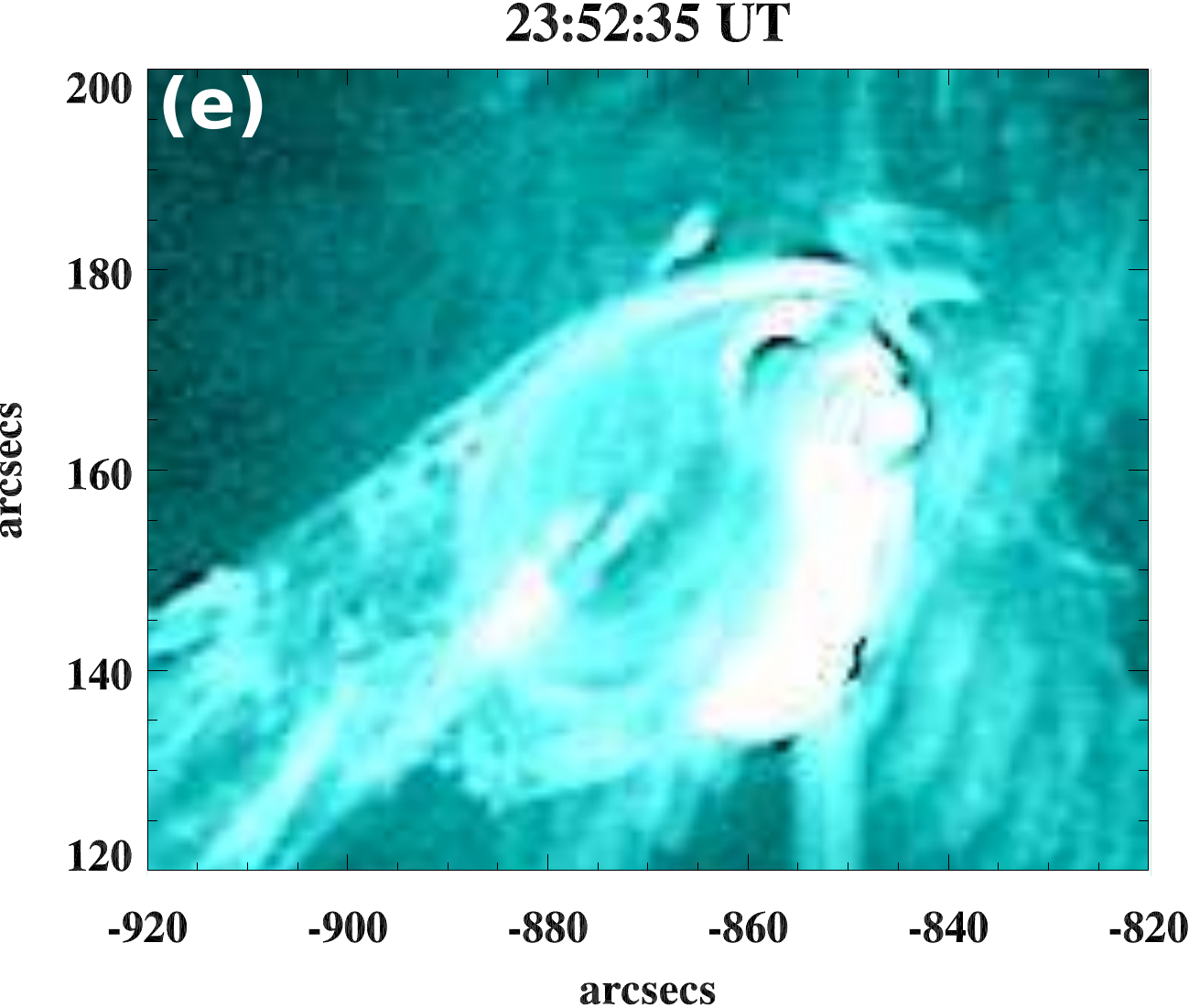}
\includegraphics[width=5.3cm]{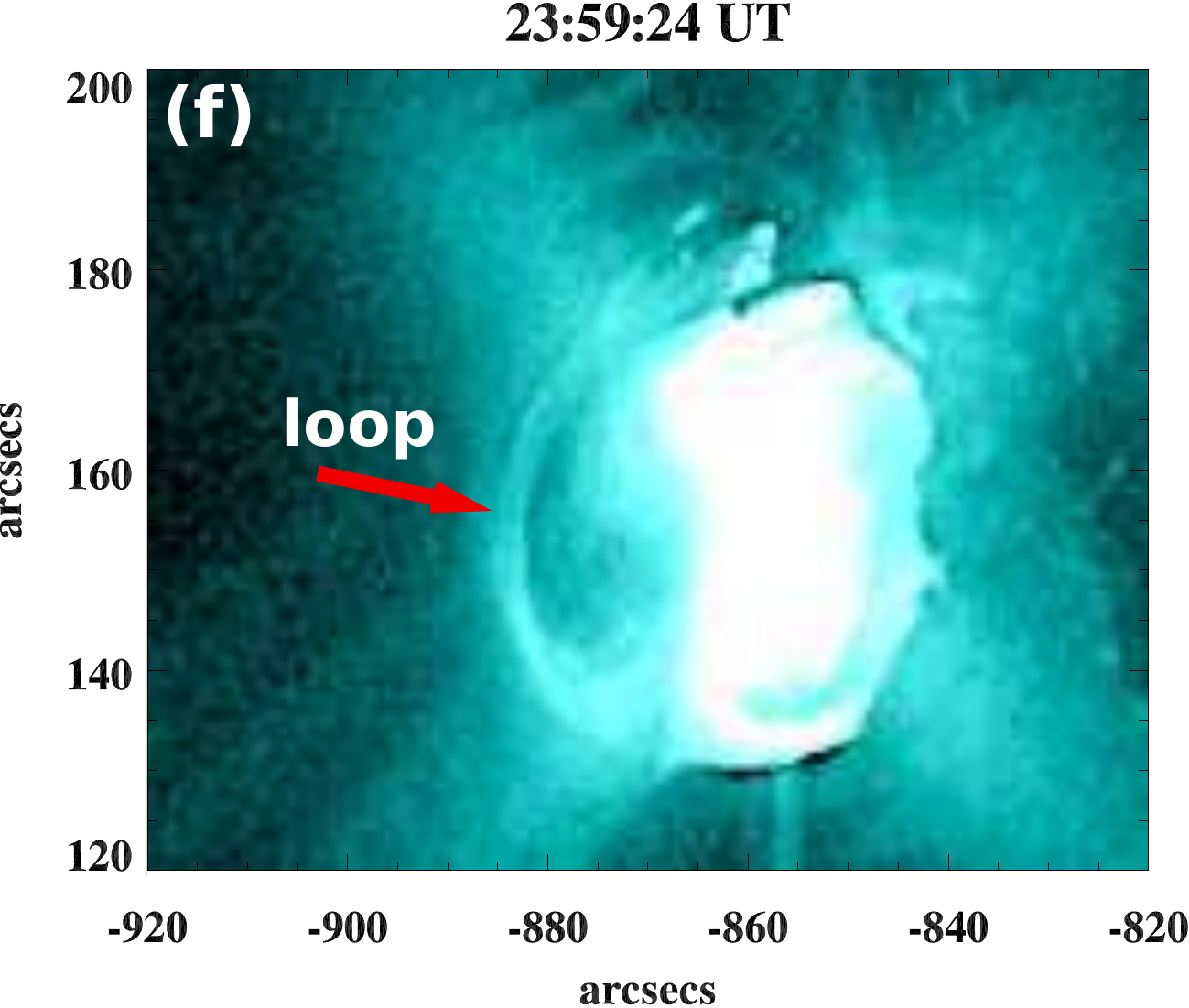}
}
\caption{{\small Selected AIA 131 \AA~ images showing the filament rise, its interaction with the pre-existing field, and associated plasma heating. (An animation of this figure is available.)}}
\label{aia131}
\end{figure*}



\begin{figure*}
\centering{
\includegraphics[width=9.0cm]{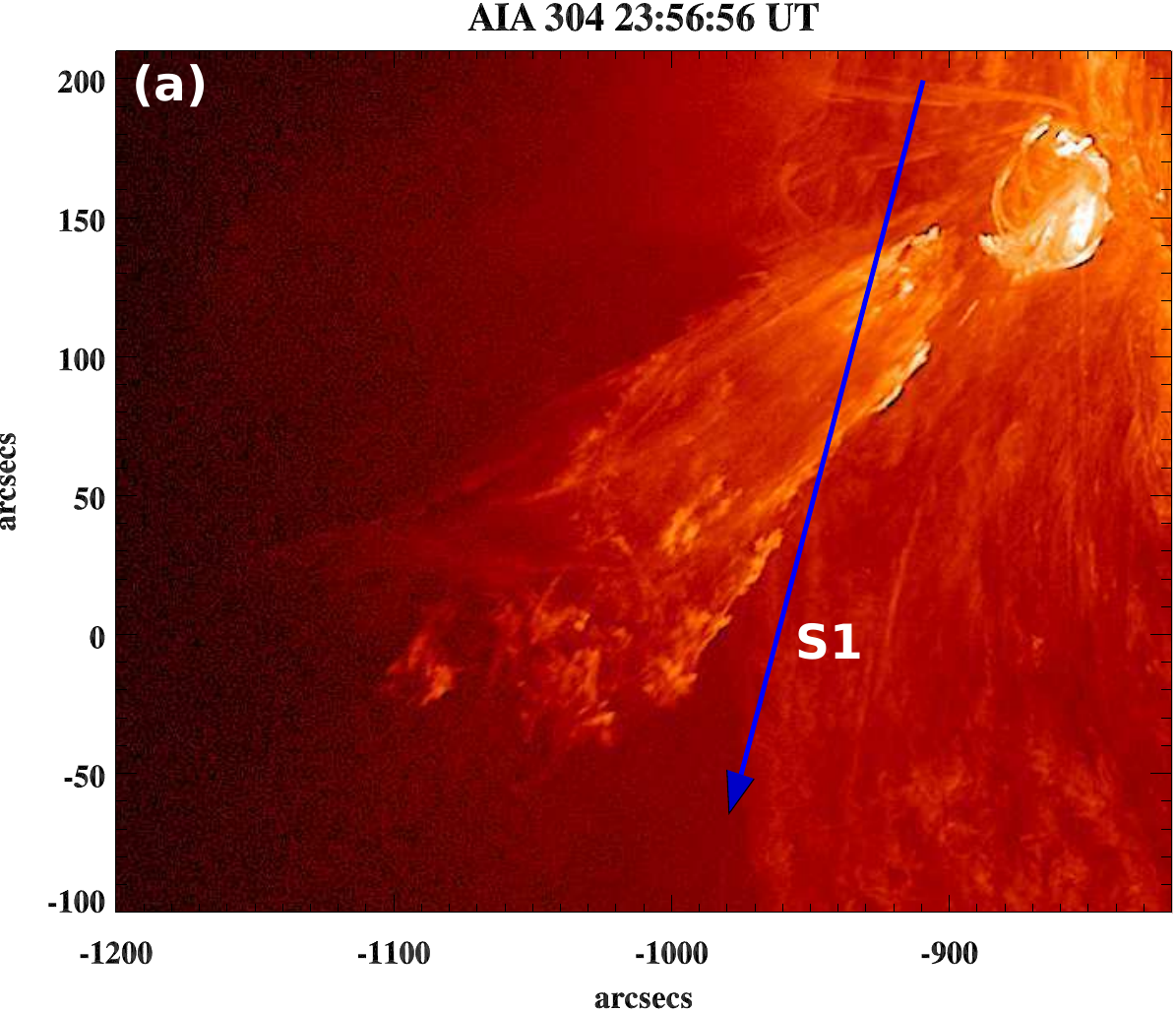}
\includegraphics[width=6.7cm]{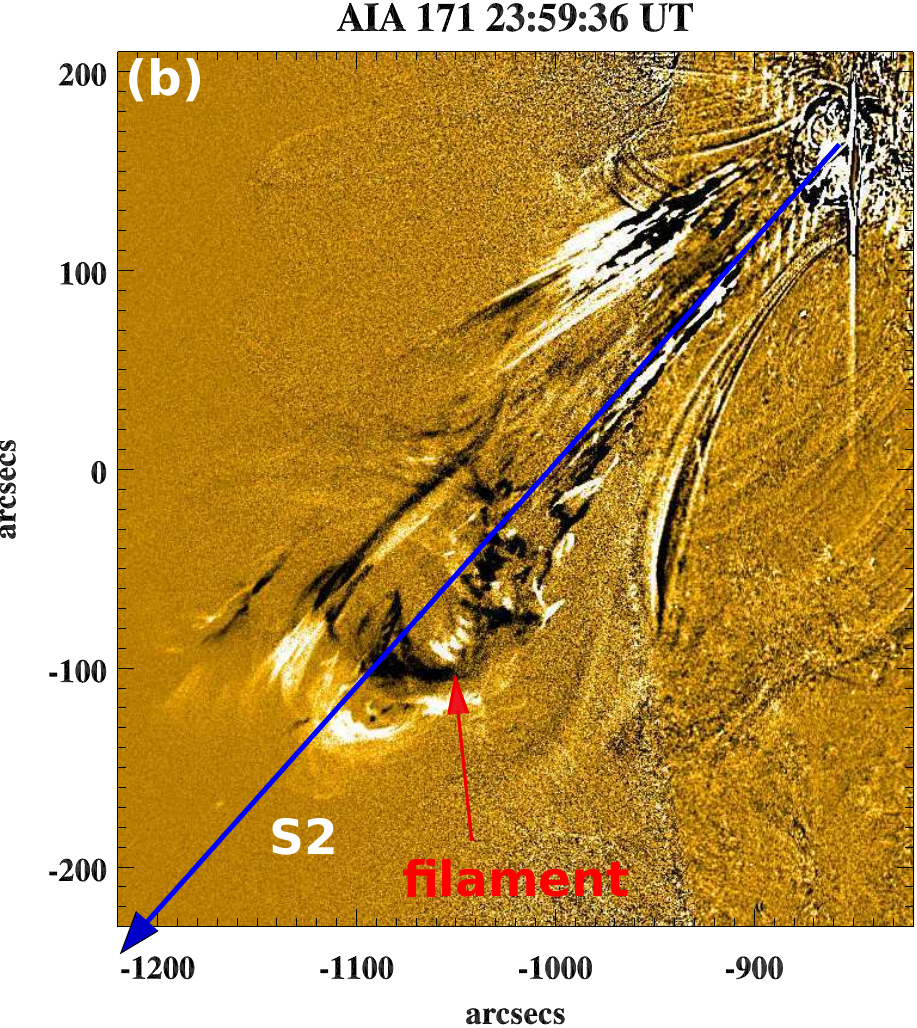}

\includegraphics[width=8.0cm]{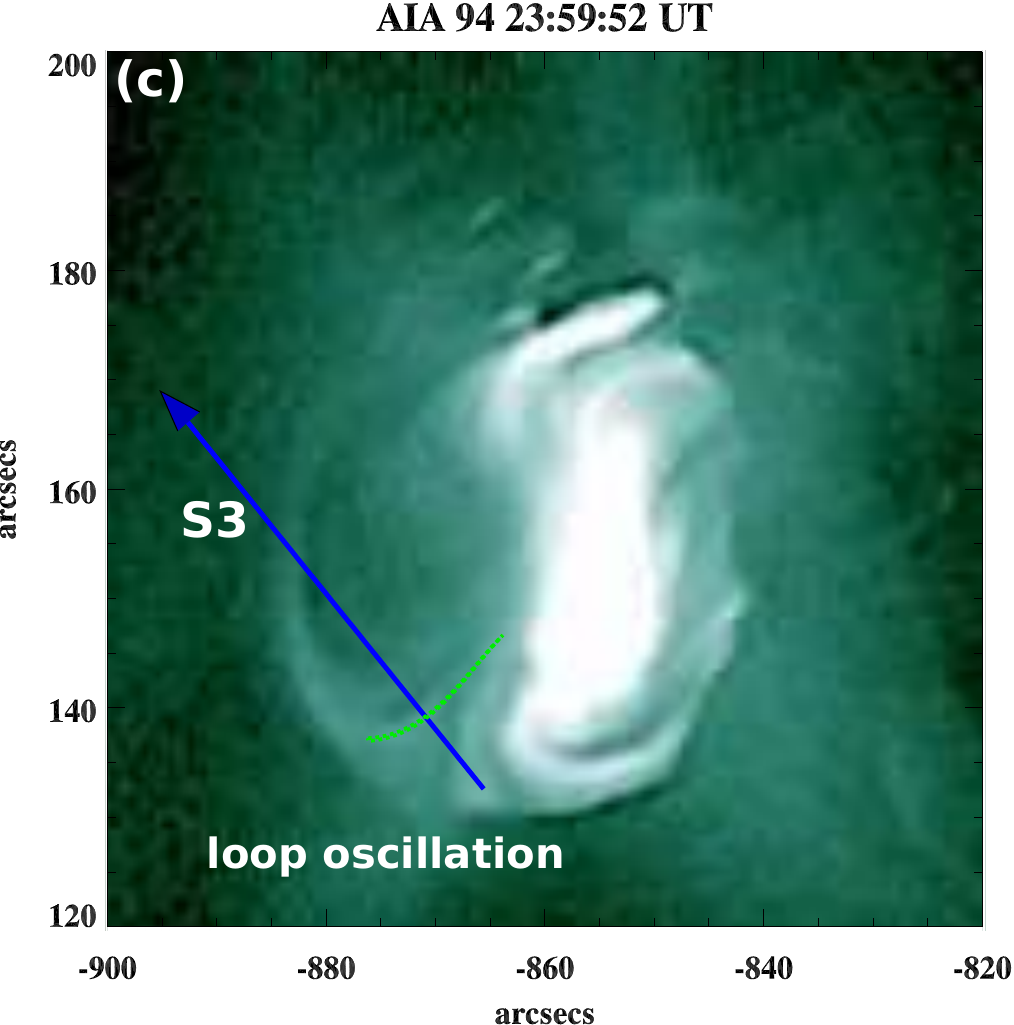}
\includegraphics[width=8.0cm]{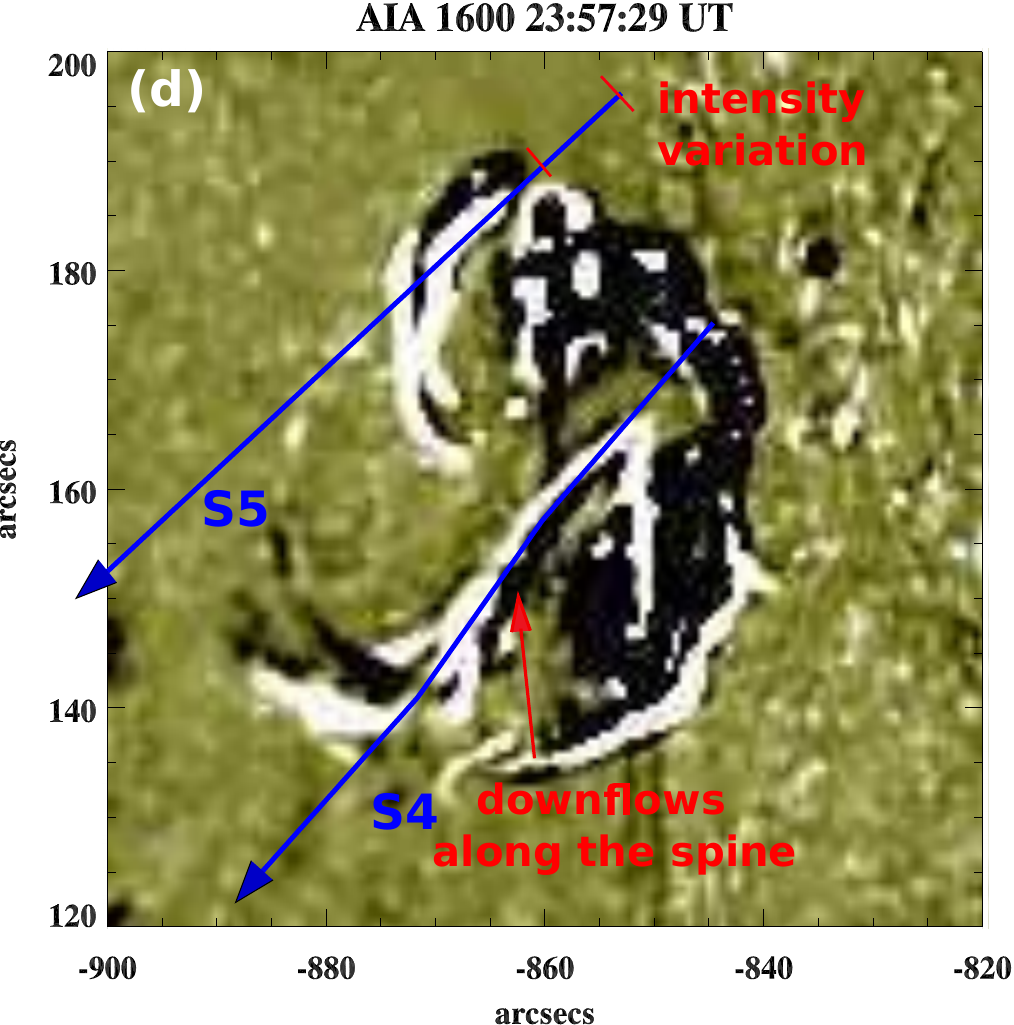}
}
\caption{{\small AIA 304, 171, 94 and 1,600 \AA~ images during the flare. S1, S2, S2, S4, and S5 are the slice cuts used to create the time-distance plots. Arrow head indicates the direction of the cut for the time-distance plots. (An animation of this figure is available.)}}
\label{aia}
\end{figure*}


\subsection{Magnetic configuration}
To investigate the magnetic configuration of the flare site, we used Hinode/SOT \citep{tsuneta2008}, XRT \citep{golub2007}, SDO/HMI and AIA images. Figure~\ref{mag}(a,b) displays HMI magnetograms of the flaring active region on 24 Sepetember (left) and 25 September 2011 (right). Since the active region was close to the eastern limb, therefore, some of the magnetic polarities at the edge of big sunspot may not be true. We carefully checked the magnetic polarities after 24 September when the active region was closer to the disc center. At the flare site, we observe negative polarities (N) surrounded by the positive polarity region (three sides). The arc shaped PIL is marked by the green line. The M-class flare occurred at the edge of the active region. The \ion{Ca}{2}~H line images (Figure~\ref{mag}(c,d)) show a quasi-circular flare ribbon. Circular or quasi-circular flare ribbons generally indicate the presence of a fan-spine topology at the flare site \citep{masson2009,pariat2010,wang2012,cheung2015,kumar2015}. We also noticed the presence of bright upflows along the possible spine (marked by the dotted line) in the chromosphere. Figure \ref{mag}(e) displays the Hinode XRT image (Be\_thin, log T$\sim$6.8-7.0 MK) during the flare onset (23:50:56 UT). Interestingly, we see quasi-circular brightening at the footpoints of the loops. The AIA 304 \AA~ image (Fig.~\ref{mag}(f)) shows a small filament lying along the neutral line. The AIA 171 \AA~ image reveals open field lines with a possible fan-spine topology at the eruption site (Fig.~\ref{mag}(g)). These images are overlaid by the HMI magnetogram contours of positive (red) and negative (blue) polarities.

The Solar TErrestrial RElations Observatory (STEREO) SECCHI-B (behind, \citealt{wuelser2004,howard2008}) observed the same active region close to the disk center. Therefore, it provides an opportunity to visualise the 3D structure of the active region loops. Also, SECCHI observed the flare and jet-like eruption of the filament material along the open field lines.
Figure~\ref{secchi}(a) displays the active region in the 171 \AA~ channel before the flare onset (at 22:14:54~UT). We can see the fine structure and connectivity highlighted by different loops within the active region. We assume fan-loops with an outer spine at the flare site. On the basis of these observations, we draw (over the image) open field lines and an outer spine at the eruption site. In Fig.~\ref{secchi}(b), we display an SECCHI 195 \AA~ image showing the flare and associated plasma flow along the outer spine at 00:06:25~UT. Similar fan loops and the plasma motion along the outer spine was observed in the SECCHI 304 \AA~ (reverse colour) image (00:07:09 UT). The schematic cartoon of the fan-spine topology including a possible 3D null point is shown in Fig.~\ref{secchi}(c).  The development of the filament resulted in eruption that followed the open field line along the outer spine and produced a CME in the interplanetary medium.

\begin{figure*}
\centering{
\includegraphics[width=7.8cm]{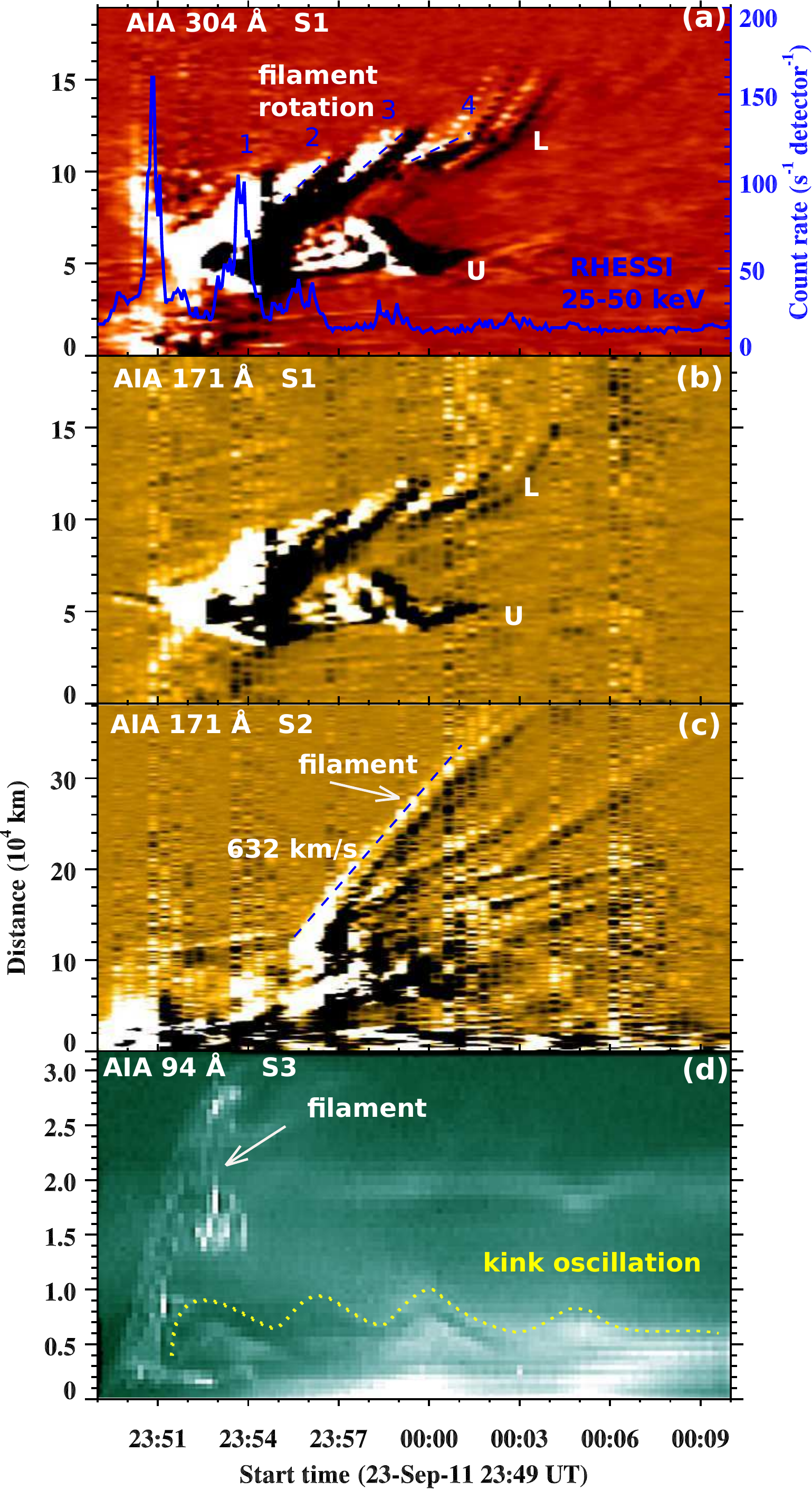}
\includegraphics[width=7.8cm]{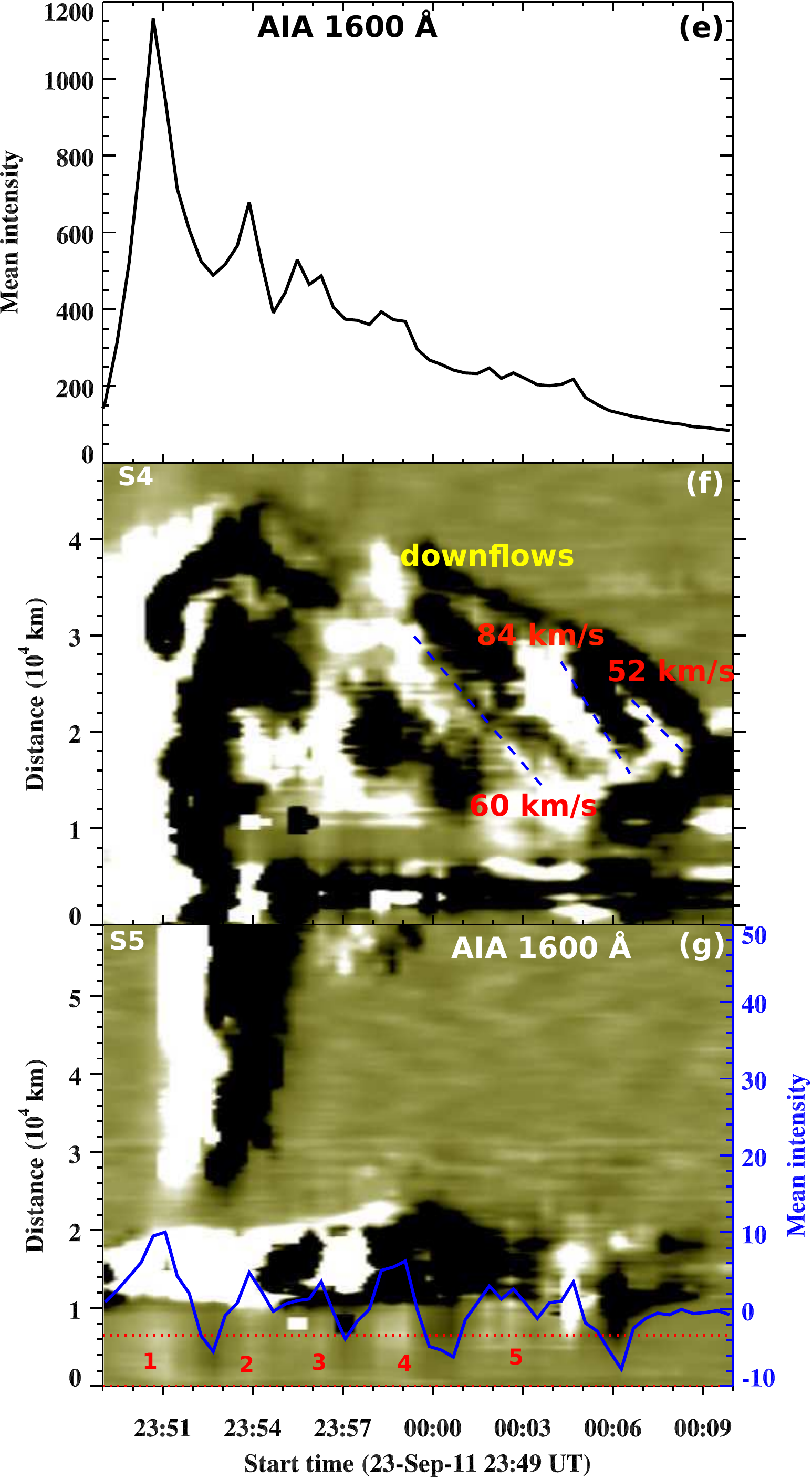}
}
\caption{{\small (a-d) Time-distance intensity plots along slices S1, S2, and S3 using AIA 304, 171, and 94 \AA~ images. AIA 304 and 171 \AA~ plots are created from the running difference images with the time of 1~min between the images;  while the AIA 94 \AA~ plot is made from the intensity images. The blue curve is the RHESSI hard X-ray flux profile in the 25--50 keV channel. U and L are the upper and lower parts of the jet structure. (e) The AIA 1,600 \AA~ mean intensity profile extracted from the box 1. (f-g) AIA 1,600 \AA~ time-distance running difference intensity plots along the slices S4 and S5. The blue curve shows the mean intensity extracted between two red horizontal dotted lines.}}
\label{stack}
\end{figure*}

\begin{figure*}
\centering{
\includegraphics[width=6.0cm]{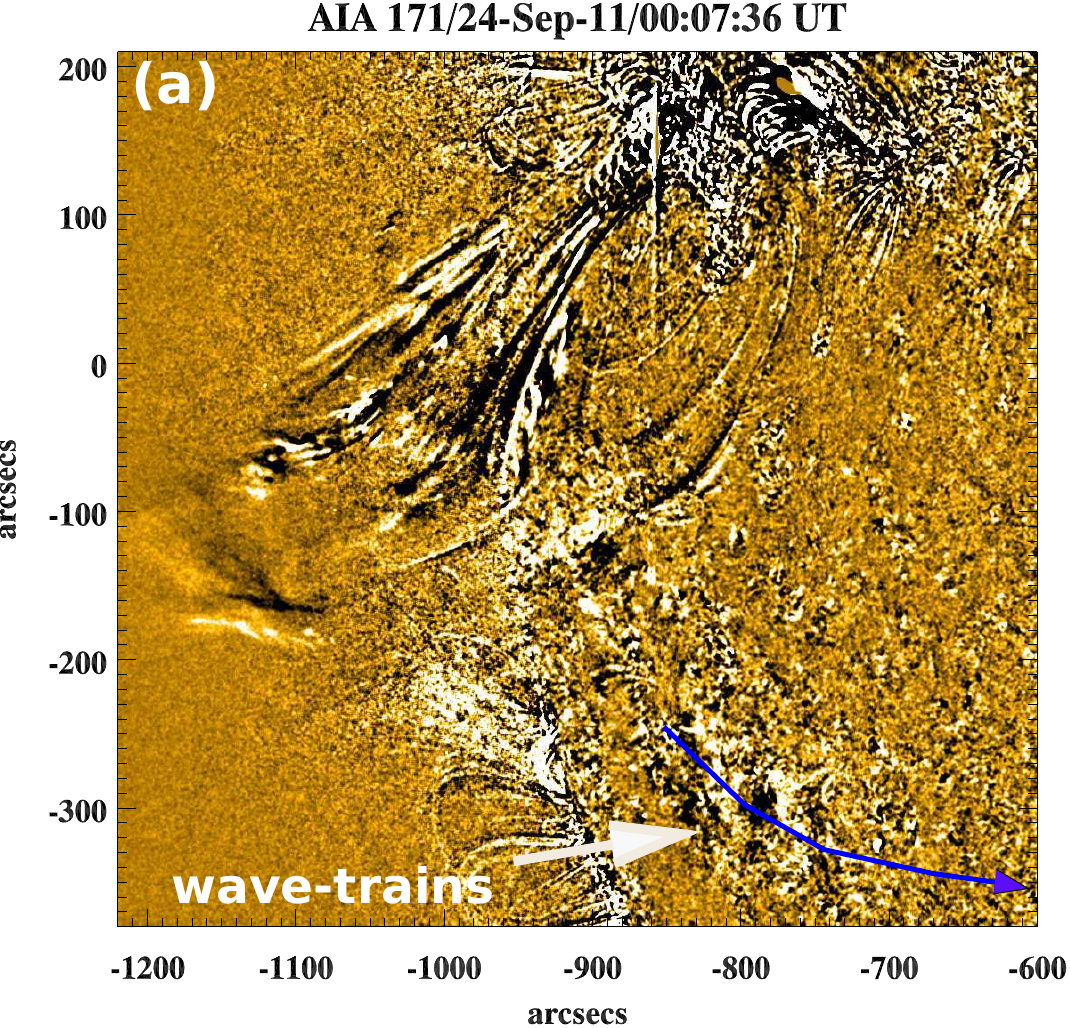}
\includegraphics[width=9cm]{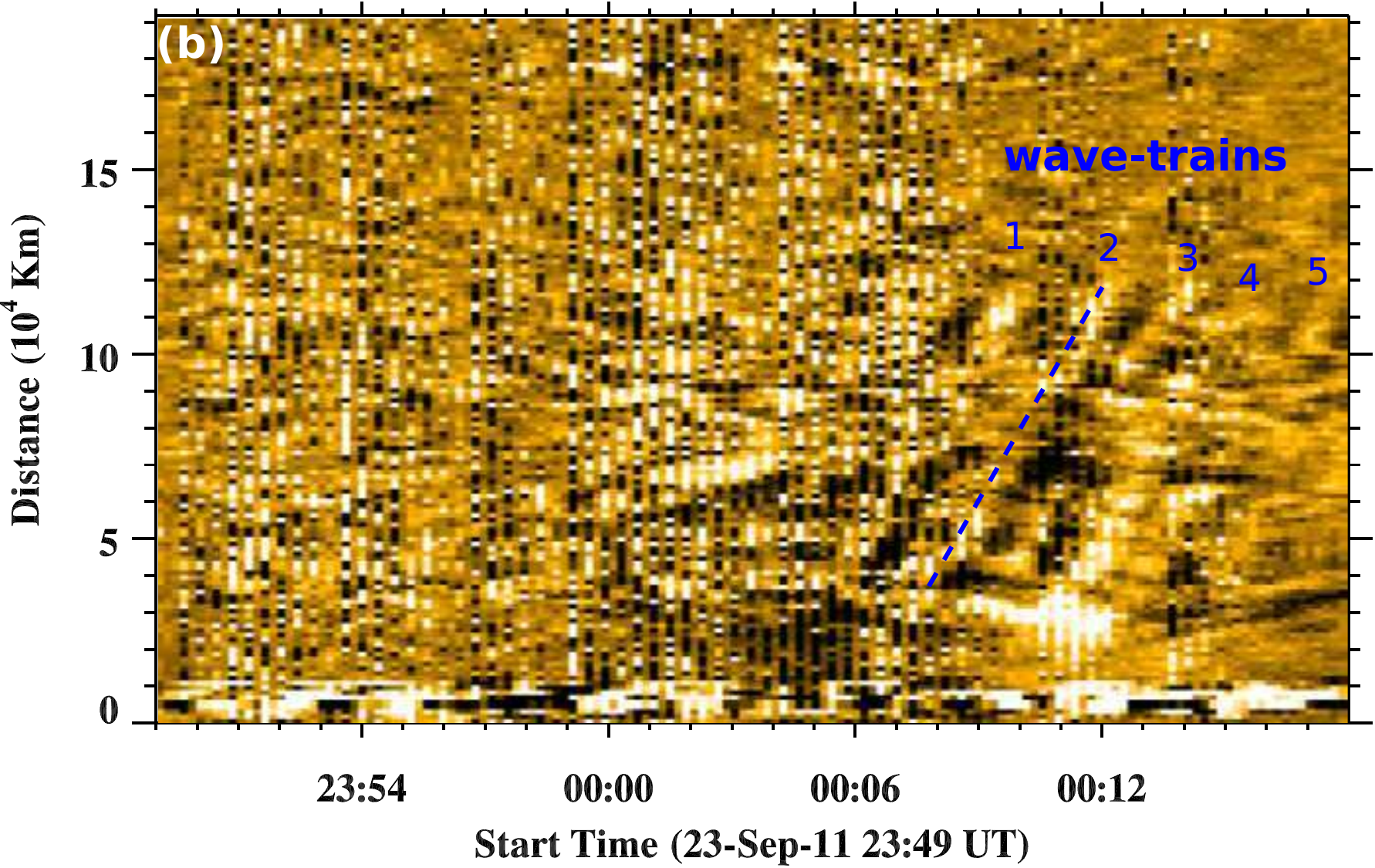}

\includegraphics[width=15cm]{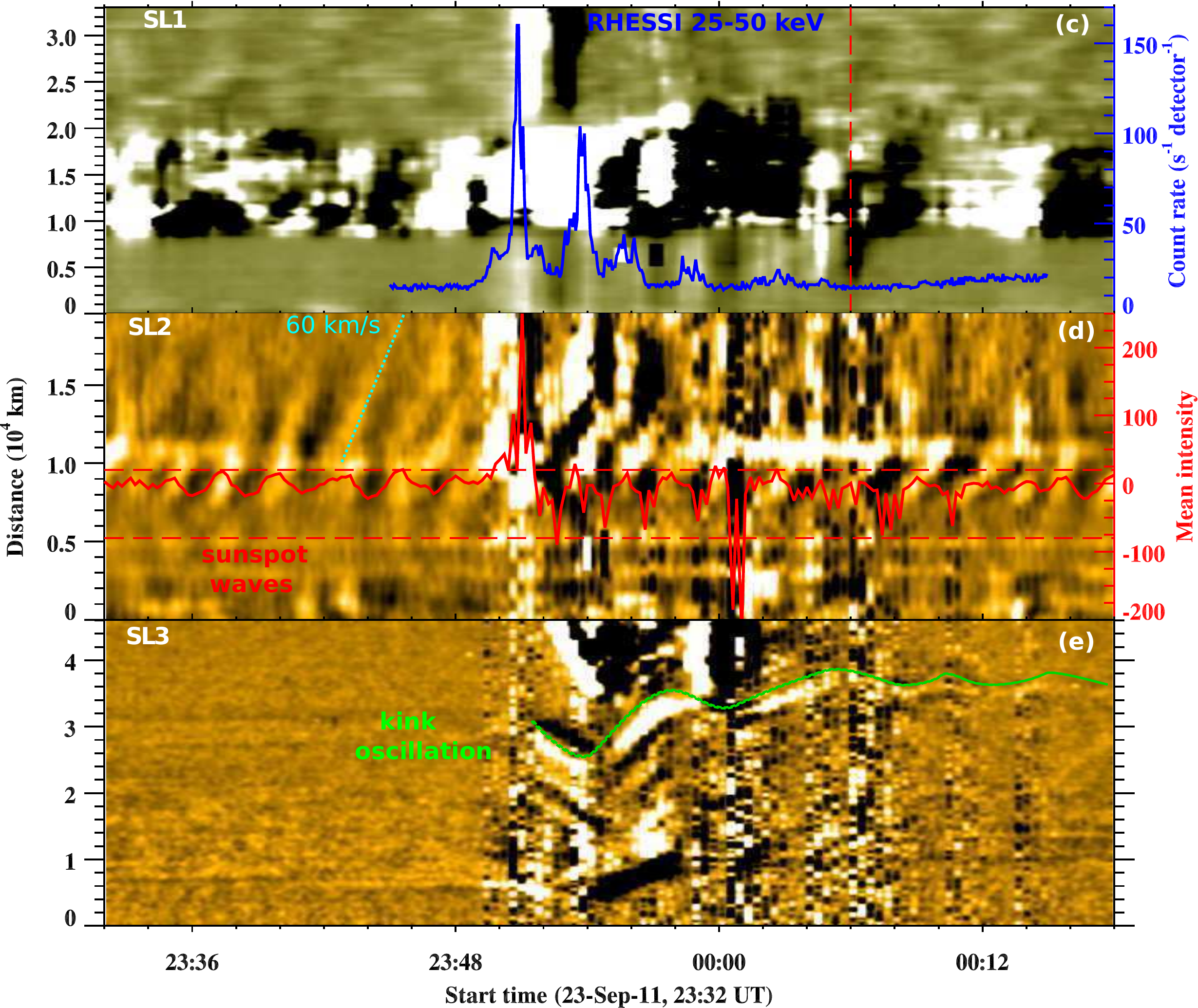}
}
\caption{{\small (a) AIA 171 \AA~ running difference image showing a rapidly-propagating wave train (b) Time-distance intensity (running difference) plot along the selected path (blue) shown in panel (a). (c) Stack plot of the AIA 1,600 \AA~ running difference intensity along slice SL1 shown in Fig.~\ref{source}(c). Blue curve is the RHESSI hard X-ray flux profile in the 25--50 keV channel. The vertical dotted line (red colour) shows the appearance of the wave train in the AIA 171 \AA~ channel. (d, e) Stack plots of the AIA 171 \AA~ running difference intensity along the slices SL2 and SL3 (refer to Fig.~\ref{mag}(f)) showing 3-min waves leaking from a sunspot and the 6-min kink oscillations (green) of the overlying arcade loops, respectively. The red curve shows the mean intensity extracted between two horizontal lines shown in the middle panel. (An animation of this figure is available.)}}
\label{train}
\end{figure*}

\subsection{Trigger of the oscillation}
Figure~\ref{aia131} displays several selected images taken in the AIA 131 \AA~ channel. The 131 \AA~ channel is sensitive to both hot ($T \approx 10$~MK) and cool plasma ($T \approx 0.4$~MK). Before the flare, we see a small filament (at $\sim$23:48~UT) lying below the fan-loops. The flare brightening starts with the rise of the filament (at 23:49:21~UT). The filament interacted with the overlying preexisting fields. The ejection of the filament is similar to the blowout jet scenario (e.g., \citealt{moore2015}). The filament material showed an untwisting motion during the eruption. We also noticed a transverse oscillation of a small loop above the flare centre (Fig.~\ref{aia131}(f)).


To study the untwisting motion of the erupting filament, its speed, the kink oscillation of the small loop, and the associated downflows along the spine at the flare site, we used slice cuts S1, S2, S3, S4, and S5 made in the AIA 304, 171, 94, and 1,600 \AA~ images (Figure~\ref{aia}). Fig.~\ref{aia}(a,b) show the filament material ejected from the flare site in the AIA 304 \AA~ intensity and 171 \AA~ running difference images. Cut S1 is used to detect the untwisting/rotation of the filament material, whereas cut S2 is used to visualise the outward plasma ejections during the periodic energy release. The direction of the slice cut is marked by the arrowhead at the end of the slice. Fig.~\ref{aia}(c) shows the AIA 94 \AA~ intensity image for the slice cut S3. Cuts S4 and S5 in the AIA 1,600 \AA~ (running difference) image are used to detect the plasma downflows along the spine (Fig.~\ref{aia}(d)).

Figure~\ref{stack}(a,b) displays the stack (running difference intensity) plots along the slice S1 using sets of the AIA 304 and 171 \AA~ images.  We see the periodic brightening in these plots, which may be associated with the untwisting motion of the filament plasma. The bright propagating features in the stack plot shows the tangential motion of the plasma. To see the temporal association of these untwisting motions with the periodic energy releases, we overplotted the RHESSI hard X-ray flux (25--50~keV, blue curve) in panel (a). Interestingly, we see a good temporal correlation of the untwisting motion with the hard X-ray bursts with a 3-min period. The periodic brightening has an approximately 3-min period similar to the hard X-ray emission.
The linear/tangential speeds of untwisting features 2, 3, and 4 (blue dotted lines) are $\sim$256, $\sim$235, and $\sim$130~km~s$^{-1}$ respectively. The average speed is $\sim$207~km~s$^{-1}$. The careful investigation of Fig.~\ref{stack}(b) reveals an additional co-temporal untwisting motions in the bottom side of the stack plots. 
It is clear that the upper part (U) of the jet shows counterclockwise motion, whereas the lower part (L) of the jet reveals the clockwise rotation. These oppositely directed untwisting motions may be a signature of a torsional Alfv\'en wave embedded in the filament material. 

Fig.~\ref{stack}(c) shows the erupting plasma motion along slice S2. The projected speed of the outward moving plasma was $\sim$632~km~s$^{-1}$. Note that there are five distinct ejections. After interacting with the overlying fields, the filament showed impulsive acceleration.

In Fig.~\ref{stack}(d), we display the stack plot along the slice S3 using 94 \AA~ intensity images. It shows that the filament rise was followed by a transverse oscillation of the small loop located above the flare centre. To see the association of the kink oscillation with the observed periodic particle acceleration, we determine the oscillation period  as $\sim$230$\pm$10~s (i.e., about 4~min). The oscillation period is not consistent with the period of the detected QPP in the energy release. Therefore, it is unlikely that this kink oscillation of the small loop could trigger the periodic reconnection. The oscillations were observed mainly in the loop leg where we noted the downflows along the vertical spine field observed in the AIA 1600 \AA. It seems that the loop oscillation is triggered by the unstable filament.

To see the quasi-periodic energy release and associated downflows in the EUV, we used AIA 1,600 \AA~ images. Fig.~\ref{stack}(e) shows the mean 1,600 \AA~ intensity obtained by averaging the signals of individual pixels in a box that covers the flare ribbon. The quasi-periodic energy release observed in this channel is quite similar to the hard X-ray emission (25--50 keV). The emission observed in the AIA 1,600 \AA~ channel is basically generated by the periodic precipitation of the nonthermal electrons propagating downward from the reconnection site. The periodic downflows are also observed along the spine (Fig.~\ref{stack}(f)). The speed of these downflows is $\sim$60, $\sim$84, and $\sim$52~km~s$^{-1}$, respectively.
The lower end of slice S5 covers a sunspot north to the flare ribbon. The running difference intensity plot clearly shows the periodic variation of the intensity  (between the horizontal red dotted lines) generated by the precipitation of energetic particles. The variation of the average intensity between the red dotted horizontal lines is overplotted in panel (g). This finding indicates that leakage of sunspot oscillations could periodically trigger magnetic reconnection.

\subsection{Wave trains}
In addition, in the AIA 171 \AA~ channel we observed a rapidly propagating wave train.
The AIA 171 \AA~ running movie shows the wave train of the intensity variation during 00:06--00:18 UT, propagating in the southward direction. To determine the apparent speed of the wave train, we chose a curved path along the propagating wave train in the AIA 171 \AA~ image (Figure \ref{train}(a)). Fig.~\ref{train}(b) displays the stack plot of running difference intensity along the selected path. We see the propagating wave train starting at 00:06 UT.  The individual wavefronts are marked by 1, 2, 3, 4, and 5. We chose the second wavefront (blue dotted line) to determine the speed, as this wave front was the most pronounced one.  The average projected speed of the front was found to be $\sim$320$\pm$27 km s$^{-1}$. The period of the filling signal in the wave train is $\sim$2.2~min.

The speed of the individual fronts in the wave train is not so high as observed in the previous flares  \citep{liu2012,kumar2013,pascoe2013}. The fast wave trains usually have higher speeds, $\sim$1000--2000~km~s$^{-1}$, measured during the impulsive phase of a flare. In our case, we observed the wave train during the flare decay phase (00:06--00:18 UT). 
 It is unclear why these wave-trains are not observed during the flare impulsive phase. However, the estimated projected speed of the wave train in our case is consistent with the result of \citet{liu2012}
who observed lateral deceleration of a wave train from $\sim$650 to 340~km~s$^{-1}$.

Let us assume that the wave train is generated during the periodic energy release. The time difference between the start time of the hard X-ray burst and the appearance of the wave trains is $\sim$15~min. The distance of the observed wave train from the flare energy release site is $\sim$300${\arcsec}$. Let us assume that the initial speed of the wave train equals to the filament material speed $\sim$632~km~s$^{-1}$. The time taken by the wave train to reach the observed distance will be $\sim$6~min. The estimated time is much smaller than the actual time difference from hard X-ray bursts. Therefore it is unlikely that the wave trains are generated during the flare impulsive energy release.

\subsection{Sunspot oscillations and kink oscillations of the 171 \AA~ arcade loops}

To investigate the relationship between sunspot oscillations and associated periodic energy release, we selected slices SL1 and SL2 (AIA 1,600 \AA~ and 171 \AA~ images shown in Fig.~\ref{source}(c) and Fig.~\ref{mag}(f)) at the sunspot located at the eruption site. The running difference intensity along these cuts (SL1, SL2) during 23:32--00:18~UT is plotted in Fig.~\ref{train}(c, d). The AIA 1,600 \AA~ movie reveals the repeated umbral flashes before the flare onset at 23:32~UT onward. Interestingly, we noticed the increase in the umbral intensity by the periodic precipitation of nonthermal electrons during the flare, which is cotemporal with the RHESSI hard X-ray flux (25--50 keV, blue curve).
Since the intensity change (marked in Fig.~\ref{aia}(d)) is consistent with the AIA 1600 mean intensity profile and HXR profile, therefore, it is likely that the intensity change could be a result of particle precipitation near the sunspot. However, we also observed the leakage of the slow-mode waves co-temporally in the AIA 171 channel (Fig.~\ref{train}(d)).

Fig.~\ref{train}(d) shows the propagating EUV disturbances, going from the sunspot along  slice SL2 (AIA 171 \AA). Their apparent (i.e., projected) speed is subsonic for the temperature associated with these EUV channels, and hence they are interpreted as slow magnetoacoustic waves \citep{demoortel2009}. In our case, the apparent speed of a front is $\sim$60 km s$^{-1}$ (Fig.~\ref{train}(d)). The average intensity (running difference) has been extracted from the 171 \AA~ stack plot (between two horizontal dashed lines) and overplotted in the same panel (red curve). A comparison of the timing of sunspot oscillations with the RHESSI hard X-ray flux profile suggests a high correlation between them. Therefore, it is very likely that the reconnection is periodically modulated by the slow-mode waves leaking from the sunspot. The possible leakage of umbral oscillations in the corona along magnetic field lines in a form of slow-mode waves has been demonstrated in \citep{botha2011}. The slow-mode waves can reach the reconnection site in the fan-spine topology and periodically trigger the energy release by the mechanism described by \citet{chen2006,sych2009}.

Furthermore, we also noticed a kink oscillation of the overlying arcade loops at the eruption site. Fig.~\ref{train}(e) shows the running difference intensity plot along slice SL3 in the AIA 171 image (Fig.~\ref{mag}(f)). During the first hard X-ray burst, we see the contraction of the arcade loops followed by decaying kink oscillations. The oscillation period is $\sim$6~min. The oscillation was possibly triggered by the erupting jet-like structure of the untwisting filament. Alternatively, the contraction of the cool arcade loops may be due to the decrease in the magnetic pressure at the flare site as a result of the magnetic energy release during the flare \citep{hudson2000}.

\subsection{Coronal Mass Ejection}
The jet-like eruption of the small filament produced a CME, which was observed during the event by the SOHO/LASCO coronagraph \citep{brueckner1995,yashiro2004}. The CME frontal loop appeared at 00:12:06~UT in the LASCO C2 field of view. The CME speed (from LASCO CME catalog\footnote[1]{\url{http://cdaw.gsfc.nasa.gov/CME_list/UNIVERSAL/2011_09/htpng/20110924.001206.p108g.htp.html}}) was $\sim$617~km~s$^{-1}$, which is close to the filament material speed ($\sim$632~km~s$^{-1}$) in the low corona (Fig.~\ref{stack}c). The CME does not show significant deceleration (2.83~m~s$^{-2}$) up to $\sim$20$R_{\odot}$ in the interplanetary medium. 

\section{SUMMARY AND DISCUSSION}
We presented multi-wavelength observations of 3-min QPP in a solar flare, detected simultaneously in the hard X-ray, radio and EUV channels. Recently, \citet{kumar2015} observed a decaying oscillation in the flaring emission, with three distinct peaks in the X-ray (6--12 keV) channel observed by the Fermi GBM (Gamma-ray Burst Monitor). Here we observed decaying oscillations in the RHESSI hard X-ray channel (25-50 keV) with five distinct peaks without any gradual broadening in the burst profile. Specific results of this study are summarised below:

(1) We observed a quasi-circular ribbon during the flare, which suggests a fan-spine magnetic topology of the flare site \citep[see][for discussion]{masson2009,pariat2010,pontin2013,kumar2015}. In \citet{masson2009}, the outer spine is connected to a remote ribbon. Non-thermal electrons accelerated at the reconnection site, propagate downwards along the fan loops and upward along the outer spine. A circular or quasi-circular ribbon is generated by the precipitation of nonthermal electrons downward along the fan loops. If the outer spine is connected within the active region (i.e., closed), we should observe a remote ribbon as a result of confined electrons along the outer spine/arcade loops that precipitate at the opposite footpoint. The formation of a remote ribbon within the active region with a fan-spine topology suggests the closed spine (e.g., \citealt{masson2009,kumar2015}). If the outer spine is open \citep{pariat2010}, we generally do not expect a remote ribbon associated with the outer spine. In addition, we should observe the type III radio bursts suggesting the injection/escape of nonthermal electrons (into the IP medium) along the open field lines. The hard X-ray sources (25-50 keV) which we observe are close to the footpoints
of the kinked filament and within/at the quasi-circular ribbon. We also observed open structures in
the AIA 171 \AA~ and STEREO images emanating from the flare site. A 3-D model of solar jets with a fan-spine topology including a magnetic null and open outer spine is proposed by \citet{pariat2010}. Our case study does not show a remote ribbon (within the AR) associated with outer spine. We only observed a quasi-circular ribbon and periodic type III radio bursts. Therefore, it indicates that the magnetic configuration (in our case) is most likely a fan-spine topology where the outer spine is open (similar to \citealt{pariat2010}).

In addition, we noted the untwisting motion of a filament material with an almost 3-min period. Therefore, the rising filament interacts with the overlying pre-existing field and can periodically trigger magnetic reconnection at the null point resulting in quasi-periodic injection of nonthermal electrons toward the footpoints of the fan loops leading to the formation of quasi-circular ribbons.

(2) The presence of periodic type III radio bursts suggests the propagation of nonthermal electrons along the open field lines. This means that the outer spine is extended into the interplanetary medium (i.e., open field lines at the eruption site) and not connected to the remote site within the active region as was observed by \citet{kumar2015}. Note that \citet{kumar2015} did not observe type III radio bursts in the interplanetary medium because of the possible confinement of the accelerated (nonthermal) electrons in the arcade loops (i.e., no escape into the interplanetary medium) within the active region. Moreover, the formation of a remote ribbon along with heating of the arcade loops (131 and 94 \AA~ channels) suggests the precipitation of nonthermal electrons toward the the opposite footpoint of the heated arcade loop, which produced a 3-min EUV QPP similar to that observed in the X-ray channel. In addition, the oscillation event studied by \citet{kumar2015} showed a failed eruption of an untwisting small filament that interacted with the fan-spine magnetic field configuration. The absence of longer hot arcade loops that could be visible in 131 and 94 \AA~ channels, in our event confirms that the particles are not accelerated/transported (along the closed spine/arcade loops) from the flaring footpoint to the opposite footpoint (i.e., remote ribbon) in the active region. Therefore, magnetic topology is open,  allowing to produce periodic type III radio bursts. In addition, in this study we observed a successful eruption of a small filament leading to a CME observed by the LASCO C2 coronagraph.

(3) This event does not reveal a slow-mode wave propagating back and forth along the hot arcade loops as reported by \citet{kumar2013w,kumar2015}. Therefore, we can rule out the trigger of periodic reconnection by a slow-mode wave reflecting back and forth between the footpoints of the hot arcade loops. In fact, the periodic reconnection may be induced by the untwisting motion of the small filament (not by a reflecting longitudinal wave in the hot arcade loops). The rising filament reconnects with the pre-existing fields, and then gets destroyed to form a jet-like structure.

(4) This study also highlights the origin of the blowout jets observed in the coronal holes, which generally reveals an untwisting motion during the eruption \citep{moore2015}. However, properties of the coronal hole blowout jets are quite similar to the active region jets originating in the fan-spine topology. This untwisting motion is generally observed during the interaction of a kink-unstable flux rope with the pre-existing ambient field. One leg of the flux rope can be detached as a result of reconnection that leads to the unwinding/untwisting motion of the flux rope \citep{kumar2012a,kumar2014,kumar2015}.


(5) The observed 3-min QPP are most likely triggered by slow-mode waves leaking form a nearby sunspot. The waves may be guided by the legs of the untwisting filament, and trigger periodic reconnection at the null point by the mechanism proposed in \citet{chen2006}. We observed a good temporal correlation between the sunspot waves and HXR QPP. A similar scenario of the leakage of sunspot oscillations in the corona, and periodic triggering (modulation) of flaring energy release was discussed in \citet{sych2009}.

(6) We observed kink oscillations of a small loop (131 and 94 \AA) at the flaring site, with a period of $\sim$4 min. In addition, we also observed kink oscillations of the overlying arcade loops (period $\sim$6 minute). The kink oscillation period does not match the period of the observed QPP ($\sim$3 minute). Therefore, we can rule out the modulation of the periodic reconnection by the kink oscillation of the nearby loops, and consider the kink oscillations as a response of the coronal fast-mode (kink) MHD resonators to the flaring energy release. However, the wave train observed in this event may be generated by the periodic magnetic reconnection.

(7) The eruption of a flux rope can launch Alfv\'en waves in the opposite direction (up and down) along the flux rope/filament and can generate a vortical motion at the footpoints in the photosphere \citep{longcope2000,fan2009}. Signatures of the torsional Alfv\'en waves have been detected in the small flux rope structures observed in the interplanetary medium \citep{gosling2010}.  
 In addition to the previous findings, our case study shows an untwisting motion of the filament material with almost the same period ($\sim$3 min) as the observed QPP. Moreover, we also noticed rotations in the opposite directions of two structures in the same jet. Very recently, \citet{lee2015} performed a numerical simulation of helical blowout jets in an emerging flux region. They found that (i) the untwisting motion had been a result of reconnection of the twisted emerging magnetic field with a pre-existing field, (ii) untwisting motion was associated with the propagation of torsional Alfv\'en waves in the corona. In addition, \citet{moore2015} reported magnetic untwisting in polar coronal hole jets and suggested the magnetic untwisting wave to be a large amplitude (nonlinear) torsional Alfv\'en wave. On the basis of the above arguments, we speculate the possibility of a torsional Alfv\'en wave in the jet-like structure. If we accept this option, it is possible that the oppositely directed Alfv\'en waves in the untwisting jet could accelerate the electrons bidirectionally with 3-min period. Alternatively, the torsional waves propagating along the flux rope could change the direction of the field lines to generate a favourable condition for periodic magnetic reconnection between the emerging twisted and pre-existing field \citep{ning2004}. The  generation of torsional waves during reconnection and associated  particle acceleration mechanism in solar flares have been proposed by \citet{fletcher2008}. 

In conclusion, we reported the possible manifestation of periodic injection of non-thermal electrons bidirectionally in a fan-spine topology during an M-class flare. The acceleration could be periodically modulated by the leakage of sunspot oscillations or by the untwisting motion of a small filament lying under the fan loops. Similar events should be investigated in more details to reveal the exact trigger mechanism of QPPs in solar flares.  

\acknowledgments
We would like to thank the referee for the positive and constructive comments that improved the manuscript considerably.
SDO is a mission for NASA Living With a Star (LWS) program. This work was supported by the \lq\lq Development of Korea Space Weather Center\rq\rq\ of KASI and the KASI basic research funds. The SDO data were (partly) provided by the Korean Data Center (KDC) for SDO in cooperation with NASA and SDO/HMI team. RHESSI is a NASA Small Explorer. Nobeyama radioheligraph and polarimeters are operated by the National Astronomical Observatory of Japan (NAOJ)/Nobeyama Solar Radio Observatory (NSRO). Hinode is a Japanese mission developed and launched by ISAS/JAXA, with NAOJ as domestic partner and NASA and UKSA as international partners. It is operated by these agencies in co-operation with ESA and NSC (Norway). PK thanks Bindu Rani for helpful discussions. K.-S. Cho acknowledges support by a grant from the US Air Force Research Laboratory, under agreement number FA 2386-14-1-4078 and by the ``Planetary system research for space exploration" from KASI.
VMN acknowledges the support from the European Research Council under the \textit{SeismoSun} Research Project No. 321141, STFC consolidated grant ST/L000733/1, and the BK21 plus program through the National Research Foundation funded by the Ministry of Education of Korea. 
Wavelet software was provided by C. Torrence and G. Compo, and is available at http://paos.colorado.edu/research/wavelets/.
\bibliographystyle{apj}
\bibliography{reference}

\end{document}